\documentclass[amsmath,amssymb,aps,superscriptaddress]{revtex4-1}
\usepackage[margin=1in]{geometry}

\usepackage{graphicx,epsfig}
\usepackage{times}
\usepackage{xcolor}
\usepackage{hyperref}
\hypersetup{
    colorlinks = true,
}
\usepackage{threeparttable}
\begin{document}

\author{Yu Liu}
\thanks{These two authors contributed equally.}
\affiliation{Department of Chemistry and Chemical Biology, Harvard University, Cambridge, Massachusetts, 02138, USA.}
\affiliation{Department of Physics, Harvard University, Cambridge, Massachusetts, 02138, USA.}
\affiliation{Harvard-MIT Center for Ultracold Atoms, Cambridge, Massachusetts, 02138, USA.}

\author{Ming-Guang Hu}
\thanks{These two authors contributed equally.}
\affiliation{Department of Chemistry and Chemical Biology, Harvard University, Cambridge, Massachusetts, 02138, USA.}
\affiliation{Department of Physics, Harvard University, Cambridge, Massachusetts, 02138, USA.}
\affiliation{Harvard-MIT Center for Ultracold Atoms, Cambridge, Massachusetts, 02138, USA.}

\author{Matthew A. Nichols}
\affiliation{Department of Chemistry and Chemical Biology, Harvard University, Cambridge, Massachusetts, 02138, USA.}
\affiliation{Department of Physics, Harvard University, Cambridge, Massachusetts, 02138, USA.}
\affiliation{Harvard-MIT Center for Ultracold Atoms, Cambridge, Massachusetts, 02138, USA.}

\author{Dongzheng Yang}
\affiliation{Institute of Theoretical and Computational Chemistry,School of Chemistry and Chemical Engineering, Nanjing University, Nanjing 210023, China}

\author{Daiqian Xie}
\affiliation{Institute of Theoretical and Computational Chemistry,School of Chemistry and Chemical Engineering, Nanjing University, Nanjing 210023, China}

\author{Hua Guo}
\affiliation{Department of Chemistry and Chemical Biology, University of New Mexico, Albuquerque, New Mexico 87131, USA}

\author{Kang-Kuen Ni}
\email[To whom correspondence should be addressed. E-mail: ]{ni@chemistry.harvard.edu}
\affiliation{Department of Chemistry and Chemical Biology, Harvard University, Cambridge, Massachusetts, 02138, USA.}
\affiliation{Department of Physics, Harvard University, Cambridge, Massachusetts, 02138, USA.}
\affiliation{Harvard-MIT Center for Ultracold Atoms, Cambridge, Massachusetts, 02138, USA.}

\title{{\Large Precision test of statistical dynamics with state-to-state ultracold chemistry}}
\date{\today}
\begin{abstract}

\begin{large}

        Chemical reactions represent a class of quantum problems that challenge both the current theoretical understanding and computational capabilities.
        Reactions that occur at ultralow temperatures provide an ideal testing ground for
        quantum chemistry and scattering theories, as they can be experimentally studied with unprecedented control, yet display dynamics that are highly complex.
        Here, we report the full product state distribution for the reaction 2KRb $\rightarrow$ K$_2$ + Rb$_2$.
        Ultracold preparation of the reactants grants complete control over their initial quantum degrees of freedom, while state-resolved, coincident detection of both products enables the measurement of scattering probabilities into all 57 allowed rotational state-pairs. Our results show an overall agreement with a state-counting model based on statistical theory,
        but also reveal several deviating state-pairs.
        In particular, we observe a strong suppression of population in the state-pair closest to the exoergicity limit, which we precisely determine to be $9.7711^{+0.0007}_{-0.0005}$ cm$^{-1}$, as a result of the long-range potential inhibiting the escape of products.
        The completeness of our measurements provides a valuable benchmark for quantum dynamics calculations beyond the current state-of-the-art.

\end{large}

\end{abstract}
\maketitle
\large

Chemical reactions, at the most fundamental level, are quantum mechanical processes where reactants are transformed into products. Consequently, a complete characterization of a reaction requires the quantum state resolution of both. Over the past decade, ultracold molecules have emerged as a powerful platform for achieving complete control over the various internal degrees of freedom of the reactants ~\cite{julienne2009ultracold,balakrishnan2016perspective,tarbutt2018laser,toscano2020cold}.
Additionally, collisions between ultracold molecules occur with the single lowest allowed partial waves ($s$- or $p$-wave) ~\cite{ospelkaus2010quantum}. 
Using these highly-controlled molecules as reactants, studies of overall reaction rates revealed the effects of long-range forces ~\cite{ni2010dipolar,guo2018dipolar,puri2019reaction} and scattering resonances ~\cite{yang2019observation, de2020imaging} with unprecedented resolution.
However, a complete characterization of these ultracold reactions at a state-to-state level has remained challenging, with progress limited to weakly-bound systems thus far ~\cite{rui2017controlled,wolf2017state}. This calls for a comprehensive method for detecting the quantum state information of the reaction products.

Much of our understanding about reactivity at the quantum level is obtained through a close interaction between experiment and theory
~\cite{yang2007state,clary2008theoretical}. Ultracold reactions bring new challenges to current reaction dynamics theories, and can play a critical role in the next stage of their development~~\cite{quemener2012ultracold}. On one hand, preparing reactants at ultralow temperatures can induce highly convoluted dynamics in reactions involving merely three or four atoms ~\cite{croft2014long}.
For example, recent studies of reactions between ultracold bialkalis revealed that the transient intermediate complexes involved can live for millions of molecular vibrations ~\cite{liu2020photo,gregory2020loss}, and exact calculations for such dynamics require computational powers beyond the state-of-the-art ~\cite{li2020advances}. On the other hand, the small sizes of these systems make them conducive to complete product quantum state mapping. Such a measurement, when combined with deterministic reactant state preparation, will provide the most precise set of benchmarks for future theories.

While the complexity of ultracold reactions hinders exact quantum calculations, statistical theories provide a viable alternative for characterizing their dynamics ~\cite{light1967statistical,nikitin2012theory,pechukas1976statistical}.
The central assumption of such theories is that the intermediate complex has sufficient time to ergodically explore the reaction phase space and redistribute its energy among the available modes of motion, leading to an equal partitioning of scattering probabilities into all allowed product channels ~\cite{bonnet1999some}.
This model has been widely used to predict the measured product state distributions of complex-forming reactions with reasonable success ~\cite{balucani2006experimental,sun2008state,rivero20111}, though systematic deviations were found and were often attributed to insufficiently long complex lifetimes.
In contrast, because of the prolonged intermediate stage of ultracold reactions,
state-to-state investigations of these systems will provide rigorous tests for statistical theories ~\cite{gonzalez2014statistical}, and allow for a critical evaluation of any non-statistical behavior ~\cite{nesbitt2012toward}. Furthermore, the precise control of the collision energy and the impact parameter in an ultracold reaction offers the possibility to examine quantum effects in product states near the energy threshold.

In this study, we investigate the product state distribution of the exchange reaction between ultracold KRb molecules prepared in their rovibronic ground state.
Using a detection scheme that combines quantum-state-selective ionization and coincidence ion imaging, we probe pairs of products (K$_2$ and Rb$_2$) that emerge from the same reaction events.
In this way, we are able to measure the scattering probabilities for all allowed product rotational state-pairs, $| N_{\textrm{K}_2}, N_{\textrm{Rb}_2}\rangle$, of which there are 57 in total. 
The resulting distribution is quantitatively compared to a state-counting model based on statistical theory through hypothesis testing. The test indicates good agreement between the measurement and the model for a subset containing $50$ state-pairs, but reveals significant deviations in several state-pairs.
In particular, a highly suppressed scattering probability is observed for the products with the lowest translational energy, which demonstrates the influence of the long-range potential on product formation.

\begin{figure} [t!] 
\centering
\includegraphics[width=1.00\textwidth]{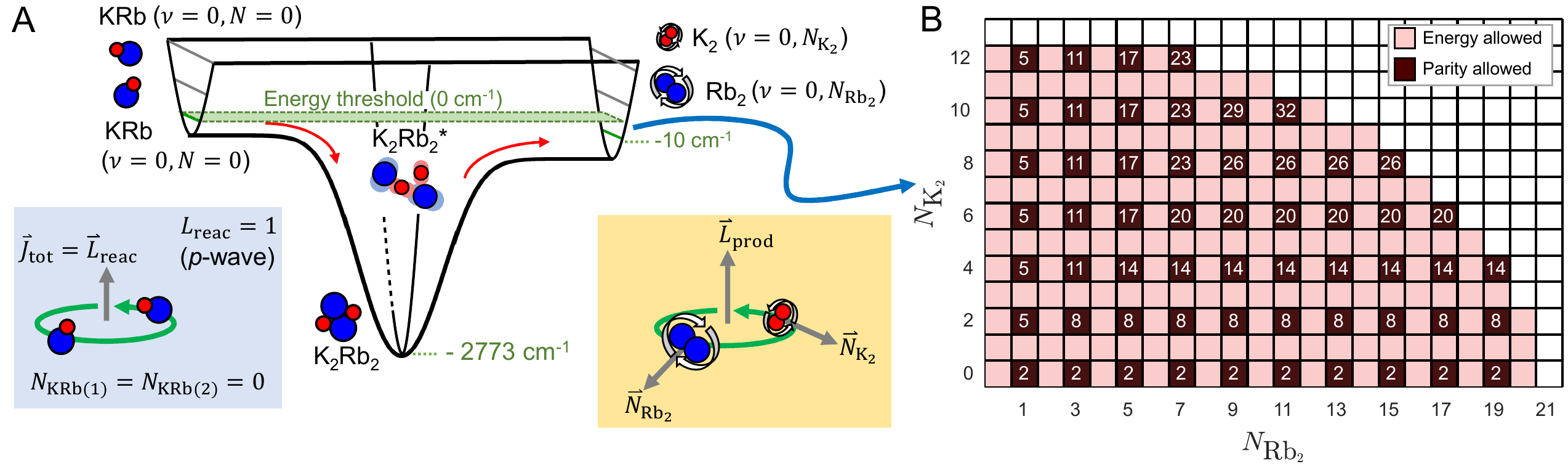}
    \caption{\normalsize\textbf{Energetics and product quantum states for ultracold reactions between KRb molecules.} \textbf{(A)} Schematic of the potential energy surface for the reaction. Reactant KRb molecules are prepared in the rovibrational ground states $|\nu = 0, N = 0 \rangle$, and the K$_2$ and Rb$_2$ products emerge in states $| \nu = 0, N_{\textrm{K}_2} \rangle$ and $| \nu = 0, N_{\textrm{Rb}_2} \rangle$. (Insets) Angular momentum vectors of the reactants and products. Two ground-state KRb molecules collide via $p-$wave collision, giving the system a total angular momentum of $J_{\textrm{tot}} = 1$. \textbf{(B)} Product rotational state-pairs and their degeneracies. The light red shaded region represents state-pairs that satisfy energy conservation, while the dark red squares represent those that additionally satisfy the parity constraint imposed by the exchange statistics of identical product nuclei. The number superimposed over each allowed state-pair represents its degeneracy, which is used to construct the state-counting model.}
\label{figSchematic}
\end{figure}

Each experiment begins with the preparation of a dilute gas of $10^4$ fermionic $^{40}$K$^{87}$Rb molecules in a single hyperfine level of their absolute ground electronic, vibrational, and rotational state~~\cite{ni2008high}. The molecules are confined within a crossed optical dipole trap (ODT), and have a peak density of $10^{12}$ cm$^{-3}$ and a temperature of 500 nK~~\cite{liu2020probing}.
Once prepared, the molecules undergo the exchange reaction~~\cite{hu2019direct}
\begin{equation}\label{KRb reaction}
\begin{split}
        \text{KRb} (\nu_{\textrm{KRb}} = 0, N_{\textrm{KRb}} = 0) + \text{KRb} (\nu_{\textrm{KRb}} = 0, N_{\textrm{KRb}} = 0) \rightarrow\ \\
        \text{K}_2\text{Rb}_2^* \rightarrow \text{K}_2 (\nu_{\textrm{K}_2} = 0, N_{\textrm{K}_2}) + \text{Rb}_2 (\nu_{\textrm{Rb}_2} = 0, N_{\textrm{Rb}_2}).
\end{split}
\end{equation}
Here, $\nu_s$ and $N_s$ are the quantum numbers associated with the vibrational and rotational degrees of freedom of species $s$, respectively. Vibrations of the products are energetically restricted to their ground states, $\nu_{\textrm{K}_2} = 0$ and $\nu_{\textrm{Rb}_2} = 0$~~\cite{hu2020product}.
The energetics of this reaction are schematically illustrated in Fig. \ref{figSchematic}A ~\cite{byrd2010structure,yang2020global}.
The deep potential well ($\sim2773$ cm$^{-1}$) and the comparatively small reaction exoergicity ($\sim10$ cm$^{-1}$), combined with the ultracold preparation of the reactants, give rise to a strong bottleneck effect for the dissociation of the K$_2$Rb$_2^*$ complex into products.
This leads to a long complex lifetime of 360~ns~~\cite{liu2020photo}, which provides a favorable condition for the complex to ergodically explore the reaction phase space and redistribute its energy equally among all available modes prior to its dissociation.

Because reaction (\ref{KRb reaction}) produces more than one molecular species, a complete characterization of its outcome requires knowledge of the population in joint quantum states of both products~~\cite{liu2007product}, which we label as $| N_{\textrm{K}_2}, N_{\textrm{Rb}_2} \rangle$.
A given state-pair is allowed if it satisfies the conservation of energy, $|\Delta E| = U(N_{\textrm{K}_2}, N_{\textrm{Rb}_2}) + T(N_{\textrm{K}_2}, N_{\textrm{Rb}_2})$, where $\Delta E$ represents the exoergicity of the reaction, while $U$ and $T$ are the internal and translational energies for the state-pair, respectively.
Given the literature value for $|\Delta E|$ ($\sim10$ cm$^{-1}$) as well as the rotational constants of K$_2$ and Rb$_2$ (Tab. \ref{tab: rovibrational constants}), energy conservation permits a total of over 200 state-pairs, as represented by the light red shaded area in Fig. \ref{figSchematic}B.
Further constraints are imposed by the exchange statistics of the identical nuclei within each product,
which restricts the allowed states to $57$ pairs wherein $N_{\textrm{K}_2}$ takes on even values and $N_{\textrm{Rb}_2}$ takes on odd values \cite{hu2020product}, as indicated by the dark red shaded squares in Fig. \ref{figSchematic}B.

Within a given state-pair, additional scattering channels arise due to the relation between the various angular momentum vectors possessed by the products. These include the rotational angular momentum of each product species, $\vec{N}_{\textrm{K}_2}$ and $\vec{N}_{\textrm{Rb}_2}$, as well as the orbital angular momentum of their relative motion, $\vec{L}_{\textrm{prod}}$ (Fig. \ref{figSchematic}A insets). Each scattering channel is associated with a unique set of orientations of these three vectors, which, under the assumption of total angular momentum conservation, must satisfy $ \vec{N}_{\textrm{K}_2} + \vec{N}_{\textrm{Rb}_2} + \vec{L}_{\textrm{prod}} = \vec{J}_{\textrm{tot}}$. Here, the quantum number for the total angular momentum of the system, $J_{\textrm{tot}}$, takes on a value of precisely 1 due to the fact that the reactant KRb molecules ($N_{\textrm{KRb}} = 0$) are fermionic and are therefore restricted to collide \textit{via} $p$-wave collisions at ultralow temperatures ~\cite{ospelkaus2010quantum}. While the scattering channels associated with a given state-pair are effectively degenerate in energy and are unresolved by our detection (SM), each represents a possible exit channel for the products, and therefore possesses an equal scattering probability under the assumption that the system behaves statistically. Thus, we construct a statistical model for the product state distribution in terms of the scattering probabilities into various state-pairs, as $P^{\textrm{0}}_{\textrm{sc}}(N_{\textrm{K}_2}, N_{\textrm{Rb}_2}) = \mathcal{D}_{N_{\textrm{K}_2}, N_{\textrm{Rb}_2}}/\sum_\mathcal{S} \mathcal{D}_{N_{\textrm{K}_2}, N_{\textrm{Rb}_2}}$, which we refer to as the state-counting model.
Here, $\mathcal{D}_{N_{\textrm{K}_2}, N_{\textrm{Rb}_2}}$ represents the number of channels associated with a given state-pair (Fig. \ref{figSchematic}B), \textit{i.e.} its degeneracy, and is counted using a set of triangle inequalities for the quantum numbers associated with the product angular momentum vectors (SM). $\mathcal{S}$ here represents the complete set of allowed state-pairs whose exact members are determined by our measurements.

\begin{figure}[t!] 
\centering
\includegraphics[width=0.70\textwidth]{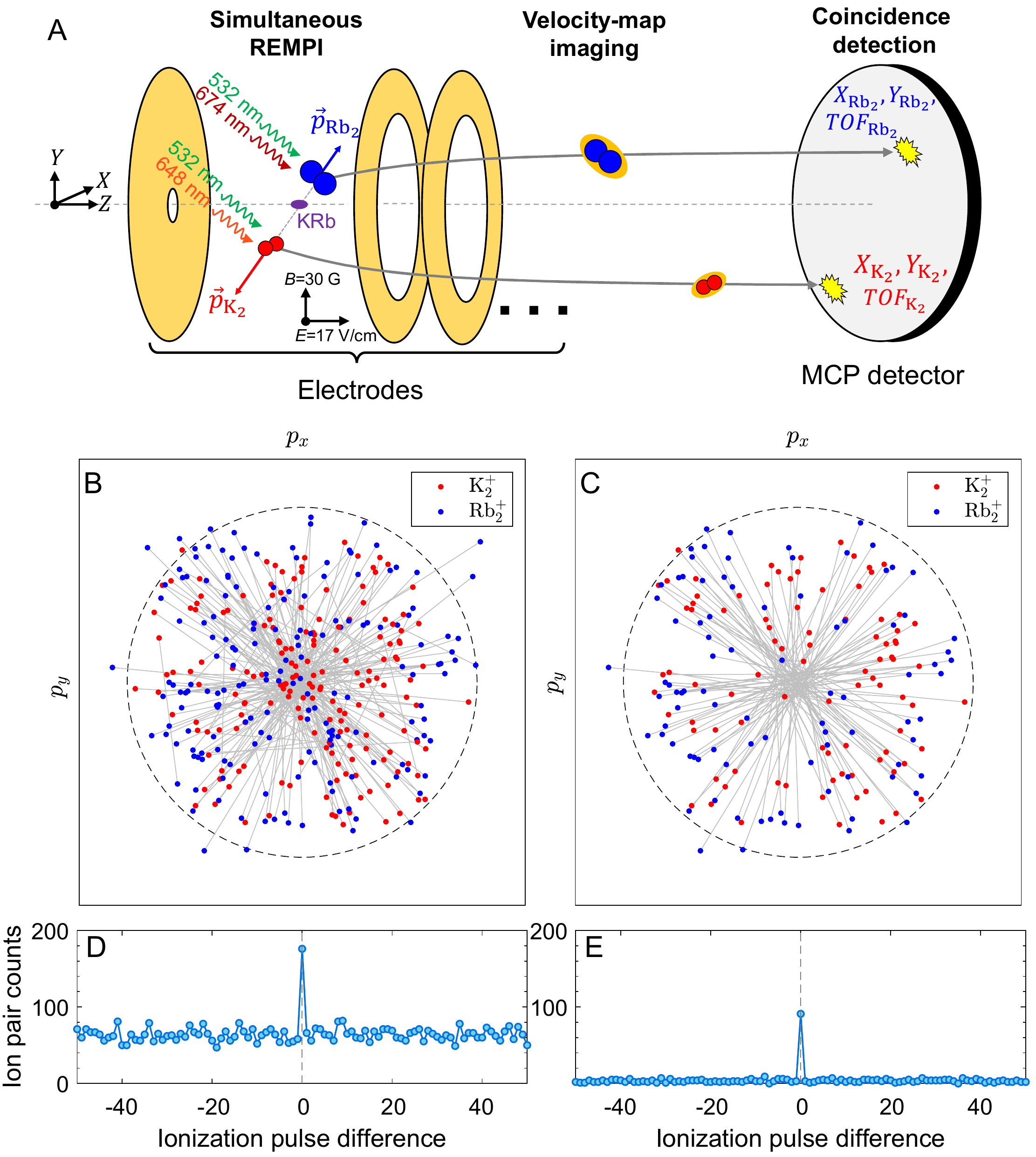}
    \caption{\normalsize\textbf{State-resolved coincidence detection of reaction products.} \textbf{(A)} Schematic of the ionization and detection process. Products in rotational states $N_{\textrm{K}_2}$ and $N_{\textrm{Rb}_2}$ are simultaneously ionized via resonance-enhanced multiphoton ionization (REMPI) pulses. The resulting ions are then velocity-map imaged onto a microchannel plate (MCP) detector. The position and TOF information of the ions are recorded and used to screen for coincidence counts. We apply a 17 V/cm electric field ($E$) for ion extraction and a 30 G magnetic field ($B$) for maintaining the nuclear spin quantization.
    \textbf{(B)} Momentum image of simultaneously detected K$_2^+$ and Rb$_2^+$ ions associated with the states $N_{\textrm{K}_2} = 6$ and $N_{\textrm{Rb}_2} = 7$. Each simultaneously detected ion pair is connected by a line. This image is derived from the raw image of ion impact positions via the position-momentum relations described in SM. The dashed circle corresponds to the maximum achievable transverse momentum for products in $|6,7 \rangle$. \textbf{(C)} Momentum image of the coincident product ion pairs, obtained by screening for the ion pairs in (B) that satisfy momentum conservation. \textbf{(D)} The number of K$_2^+$--Rb$_2^+$ pairs as a function of the number of ionization pulses by which their detections are separated. Zero on the horizontal axis corresponds to simultaneous counts, which contains both correlated (coincident) and uncorrelated (accidental) ions, while a non-zero difference correspond to ions generated by separate pulses, and are therefore uncorrelated. \textbf{(E)} A plot similar to (D), but with the uncorrelated counts screened away.}
\label{figDetectionScheme}
\end{figure}

To probe the population in a given product state-pair, we developed a state-resolved coincidence detection scheme (Fig. \ref{figDetectionScheme}A),
which involves three main steps: simultaneous state-selective ionization of K$_2$ and Rb$_2$ \textit{via} laser pulses, velocity-map imaging (VMI) of the resulting ions, 
and determination of the number of K$_2^+$ and Rb$_2^+$ ion pairs which are associated with the same reaction events. We focus our discussions here on the third step, as details of the first two steps can be found in Refs. ~\cite{liu2020probing, hu2020product} and the SM.
Each simultaneous observation of a K$_2^+$ ion and a Rb$_2^+$ ion represents a possible detection of products generated by the same event in the target state-pair $| N_{\textrm{K}_2}, N_{\textrm{Rb}_2}\rangle$ -- a coincident count. Such an observation, however, could also be due to products generated by separate reactions
-- an accidental count.
In order to identify the coincident counts, we utilize the fact that products from the same event must satisfy the conservation of linear momentum, $\vec{p}_{\textrm{K}_2} + \vec{p}_{\textrm{Rb}_2} = 0$,
while those that are from separate events are uncorrelated, and are therefore not bound by this constraint.
In our system, the momentum components transverse to the TOF axis are mapped to spatial positions on an ion detector through VMI, while the component along this axis is encoded in the ion TOF (SM).
To illustrate this screening process, Fig. \ref{figDetectionScheme}B shows an image of all simultaneous ion pairs for the state-pair $| 6, 7\rangle$,
while Fig. \ref{figDetectionScheme}C highlights those that satisfy momentum conservation and are considered to be from the same reaction events.
We assess the efficacy of this process by applying it to detected K$_2^+$ and Rb$_2^+$ that are ionized by different laser pulses, and must therefore be uncorrelated (Fig. \ref{figIonTiming} and SM).
Fig. \ref{figDetectionScheme}D and \ref{figDetectionScheme}E display the number of K$_2^+$--Rb$_2^+$ pairs before and after screening, respectively, as a function of the difference in the pulse number,
demonstrating that the uncorrelated counts are effectively screened away.

\begin{figure} [t!]
\centering
\includegraphics[width=1.00\textwidth]{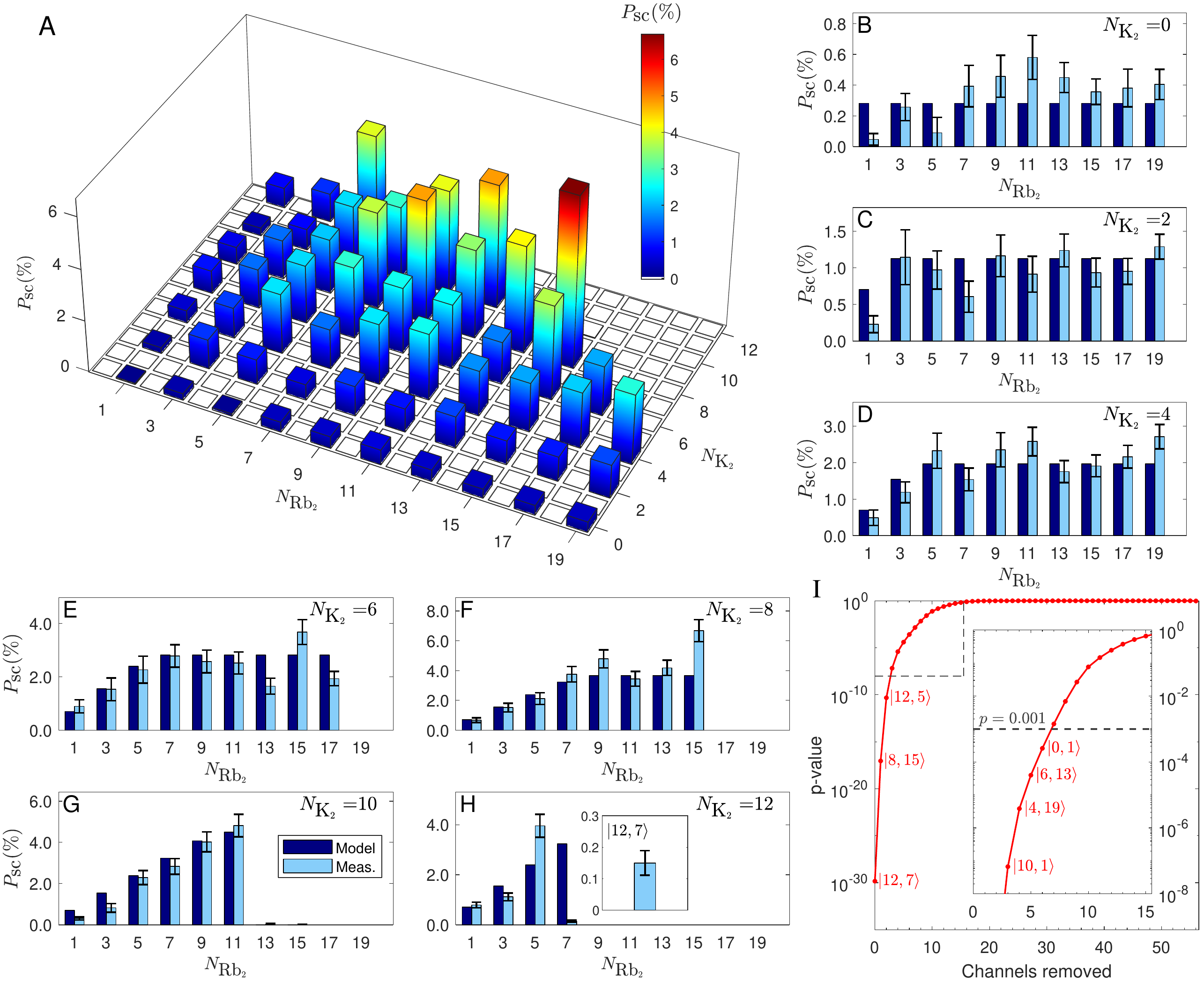}
    \caption{\normalsize\textbf{Measured product state distribution and comparison to statistical theory.} \textbf{(A)} Measured scattering probability into various product state-pairs $| N_{{\textrm{K}}_2}, N_{{\textrm{Rb}}_2}\rangle$. \textbf{(B -- H)} Comparing the measured ($P_{\textrm{sc}}^{\textrm{meas}}$) scattering probabilities to predictions ($P_{\textrm{sc}}^{\textrm{0}}$) from the state-counting model. Each frame displays the probabilities for state-pairs with a particular value of $N_{{\textrm{K}}_2}$, and as a function their $N_{{\textrm{Rb}}_2}$. The error bar for each measurement includes shot noise as well as 11\% relative fluctuations in experimental conditions. The state-pairs $| 10, 13 \rangle$ and $| 10, 15 \rangle$ are energetically forbidden from being populated, and have measured populations that are consistent with zero. (Inset of (H)) A close up view of the scattering probability for the state-pair $| 12, 7 \rangle$, which displays a strongly suppressed population compared to the prediction. \textbf{(I)} $p$-value for the hypothesis that the measured and model distributions agree, as state-pairs that deviate most significantly from prediction (labled in the figure) are successively removed. An 11$\%$ relative fluctuation in experimental conditions was considered for this calculation (SM). (Inset) A zoom-in over the boxed region. The dashed line indicates $p = 0.001$, a threshold below which the hypothesis should be rejected.}
\label{figProdStateDist}
\end{figure}

Using this coincidence detection scheme, we observe coincidence signals for all state-pairs with internal energies less than or equal to that of $|12, 7 \rangle$ ($U = 9.77$ cm$^{-1}$). 
The next state pair that is higher in energy, $| 10, 13 \rangle$ ($U = 10.01$ cm$^{-1}$), shows a signal that is consistent with zero.
This allows us to determine the complete set of allowed state-pairs, $\mathcal{S}$, which contains 57 total members.
Fig. \ref{figProdStateDist}A shows the measured product state distribution, defined as the observed probabilities to scatter into the various state-pairs, $P^{\textrm{meas}}_{\textrm{sc}}(N_{\textrm{K}_2}, N_{\textrm{Rb}_2}) = \mathcal{N}_{N_{\textrm{K}_2}, N_{\textrm{Rb}_2}}/\sum_\mathcal{S} \mathcal{N}_{N_{\textrm{K}_2}, N_{\textrm{Rb}_2}}$. Here, $\mathcal{N}_{N_{\textrm{K}_2}, N_{\textrm{Rb}_2}}$ represents the normalized coincident counts for a given state-pair, which are obtained through a normalization of the raw coincident counts by the number of experimental cycles, fluctuations in experimental conditions, and the product-velocity-dependent efficiencies of our ionization sampling (SM).
The results demonstrate, in general, enhanced scattering probability for state-pairs which have both large and closely matching values of $N_{\textrm{K}_2}$ and $N_{\textrm{Rb}_2}$ (\textit{e.g.} $| 10, 11 \rangle$), while scattering into state-pairs with small $N_{\textrm{K}_2}$ or $N_{\textrm{Rb}_2}$ is disfavored.

We compare the measured product state distribution to the state-counting model ($P_{\textrm{sc}}^{\textrm{0}}$) in Fig. \ref{figProdStateDist}B-H.
Each measurement is assigned an error bar of $\pm \delta P_{\textrm{sc}}^{\textrm{meas}}$, which represents the measurement uncertainty that arises primarily from the Poissonian statistics of the coincidence ion counting as well as fluctuations in experimental conditions (SM).
We quantify the degree to which the measured and predicted distributions agree using the likelihood ratio test ~\cite{wasserman2013all}. Specifically, we test the hypothesis that the observation matches our model, as state-pairs that deviate most significantly from the predictions are successively removed from the test (SM). For each new subset of state-pairs, the $p$-value for the hypothesis is calculated to characterize its statistical significance. The results, displayed in Fig. \ref{figProdStateDist}I, show that the threshold of $p=0.001$, above which the hypothesis cannot be rejected, is reached after the removal of 7 state-pairs (labeled in the figure). Thus, for a subset that contains the majority of the allowed state-pairs, we find that the measured outcome to be consistent with the model.
Since all reactants are prepared in a single quantum state, such an outcome cannot be attributed any ensemble averaging effect, but must be due to intrinsic dynamics of the reaction.

\begin{figure} [t!]
\centering
\includegraphics[width=\textwidth]{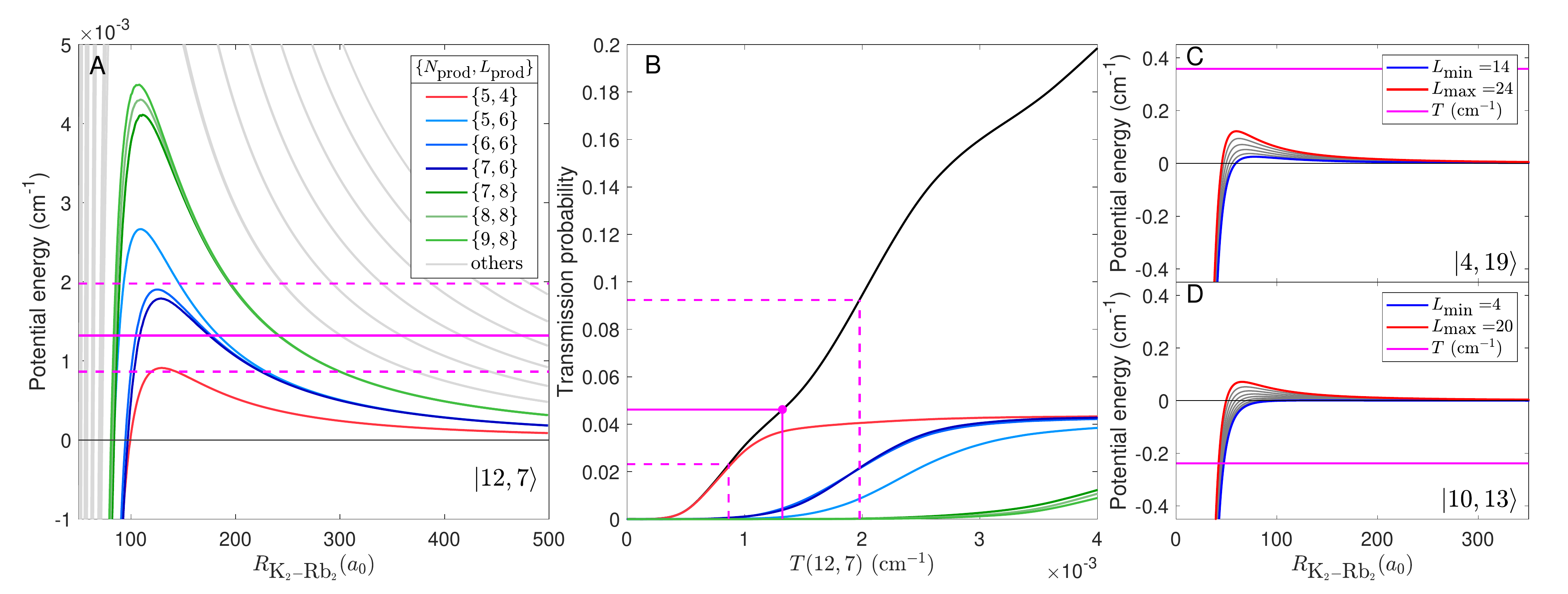}
    \caption{\normalsize\textbf{Influence of the long-range potential on product formation near the energy threshold}. \textbf{(A)} \textit{Ab initio} effective long-range potentials for different scattering channels belonging to the state-pair $|12,7\rangle$. Each channel is defined by the quantum numbers $N_{\textrm{prod}}$ and $L_{\textrm{prod}}$, associated with the coupled rotational ($\vec{N}_{\textrm{prod}} = \vec{N}_{\textrm{K}_2} + \vec{N}_{\textrm{Rb}_2}$) and orbital ($\vec{L}_{\textrm{prod}}$) angular momenta of the products, respectively. \textbf{(B)} Total transmission probability (black curve) through the centrifugal barriers and its contributions from different channels (curves of other colors) as functions of the translational energy of $|12,7\rangle$. The color-coding scheme follows that of (A). The maximum contribution of each channel is given by the inverse of the degeneracy of this state-pair, $1/23 = 0.435 $, and is reached when the translational energy increases sufficiently above the corresponding barrier height. We use the ratio between the measured and predicted scattering probabilities (Fig. \ref{figProdStateDist}H) of 0.046 as an estimate for the transmission probability, and allow variations of $+0.046$ and $-0.023$ to account for uncertainties in the population prior to barrier transmission. This corresponds to a translational energy of $1.3_{-0.5}^{+0.7}\times 10^{-3}$ cm$^{-1}$ (magenta lines). This value is also indicated in (A). \textbf{(C,D)} Potentials for the state-pairs $|4, 19\rangle$ and $|10, 13\rangle$. The magenta line in each sub-figure indicates the translational energy of the corresponding state-pair, which is calculated based on the reaction exoergicity determined in this study. $L_{\textrm{min}}$ and $L_{\textrm{max}}$ respectively represent the minimum and maximum allowed orbital angular momentum for the state-pair.}
\label{figExitChan}
\end{figure}

Because of the precise control over the collision energy, our experiment is sensitive to effects of the long-range potential on product formation in near-threshold states.
In particular, we observe a highly suppressed scattering probability into the state-pair $|12,7\rangle$ -- the measured scattering probability is only $4.6\%$ of the prediction -- which we attribute to centrifugal barriers impeding the formation of low translational energy products in this state-pair. Such an effect is beyond the state-counting model, which implicitly assumes a unit probability for product escape.
To characterize this effect, we calculated the effective long-range potentials associated with the different scattering channels of $| 12, 7 \rangle$ (Fig. \ref{figExitChan}A), as well as the probability for products to transmit through the associated centrifugal barriers as a function of the total translational energy, $T(12,7)$ (Fig. \ref{figExitChan}B) (SM). 
We use $0.046^{+0.046}_{-0.023}$ as the experimentally measured transmission probability, which assumes that the population in $|12, 7\rangle$ prior to transmission is between twice and half of that given by the state-counting model.
Comparing this value against the curve in Fig. \ref{figExitChan}B, we find the translational energy of this state-pair to be $T(12,7) = 1.3_{-0.5}^{+0.7}\times10^{-3}$ cm$^{-1}$. At such an energy, the orbital motion of the products is predominantly characterized by the single lowest allowed partial wave for $| 12, 7 \rangle$, the $g$-wave ($L_{\textrm{prod}} = 4$). This minute energy scale could enable the control of product formation \textit{via} external fields ~\cite{gonzalez2014statistical, meyer2010product}, thereby extending the extraodinary controllability over ultracold reactants to products.

Using the newly determined translational energy and the known internal energy of $| 12, 7 \rangle$, we calculate the reaction exoergicity to be $|\Delta E| = T(12,7) + U(12,7) = 9.7711^{+0.0007}_{-0.0005}$ cm$^{-1}$. To the best of our knowledge, this represents the most precise determination of an exoergicity for any bimolecular chemical reaction. Based on this value, we calculate the translational energies of two state-pairs, $|4, 19\rangle$ and $|10, 13\rangle$, which lie just above and below the energy threshold, respectively. The results are displayed along with the respective sets of effective potentials in Figs. \ref{figExitChan}C and D. Here, we see that products formed in $|4, 19\rangle$ escape with energy far above all centrifugal barriers, while those formed in $|10, 13\rangle$ have insufficient energy to escape the complex, consistent with our measurements.

The breakdown of the state-counting model for the most near-threshold state-pair highlights the importance of considering the escape process of the products. To this end, we calculate the product escape probabilities for the remaining 56 state-pairs by explicitly following their dynamics over the long-range potential ~\cite{yang2020statistical} (SM). The results show near unit probabilities ($>$0.999) for all 56, indicating a lack of any barriers or bottlenecks that impede product formation in these states. This also implies that the deviations observed in other state-pairs (\textit{e.g.} $|8,15\rangle$ and $|12,5\rangle$) may originate from non-statistical dynamics ~\cite {nesbitt2012toward}, a definitive explanation for which will require exact quantum scattering calculations beyond the current state-of-the art.

While the reactants used in the present study were prepared in their absolute ground states, our molecular state control can be readily extended to allow preparation in arbitrary rotational and vibrational states, or even superposition states with controllable relative phases. By combining the ability to measure product quantum state information in a pair-correlated fashion, as we have demonstrated here, ultracold reactions represent a promising platform to study quantum effects such as geometric phase~~\cite{kendrick2015geometric,kendrick2020non}, interference~~\cite{brumer2000identical}, and entanglement~~\cite{molina2015quantum} in chemical reactions with unprecedented precision.

\clearpage

\noindent \textbf{Acknowledgements} We thank L. Zhu for experimental assistance; T. Rosenband, G. Qu\'{e}m\'{e}ner, W. Cairncross, E. Heller, and M. Soley for insightful discussions; T. Karman for providing the code for state-counting; and L. Liu for a critical reading of the manuscript. This work is supported by the DOE Young Investigator Program (DE-SC0019020) and the David and Lucile Packard Foundation. M.A.N. is supported by the Arnold O. Beckman Postdoctoral Fellowship in Chemical Instrumentation. D.Y. and D.X. acknowledge support from National Natural Science Foundation of China (Grant Nos. 21733006). H.G. thanks the Army Research Office (W911NF-19-1-0283) for funding.

\noindent \textbf{Competing interests:} The authors declare that they have no competing financial interests. 

\noindent \textbf{Data and materials availability:} Data from the main text and supplementary materials are available from the corresponding author upon reasonable request.

\clearpage

\section*{\Large Supplementary materials}

\setcounter{table}{0}
\renewcommand{\thetable}{S\arabic{table}}
\renewcommand{\thefigure}{S\arabic{figure}}
\setcounter{figure}{0}
\renewcommand{\thesection}{\large S\arabic{section}}
\renewcommand{\theequation}{S.\arabic{equation}}

\section{\large~Experiment timing during ionization} \label{section: timing}

\begin{figure}[t!]
\centering
\includegraphics[width=0.75\textwidth]{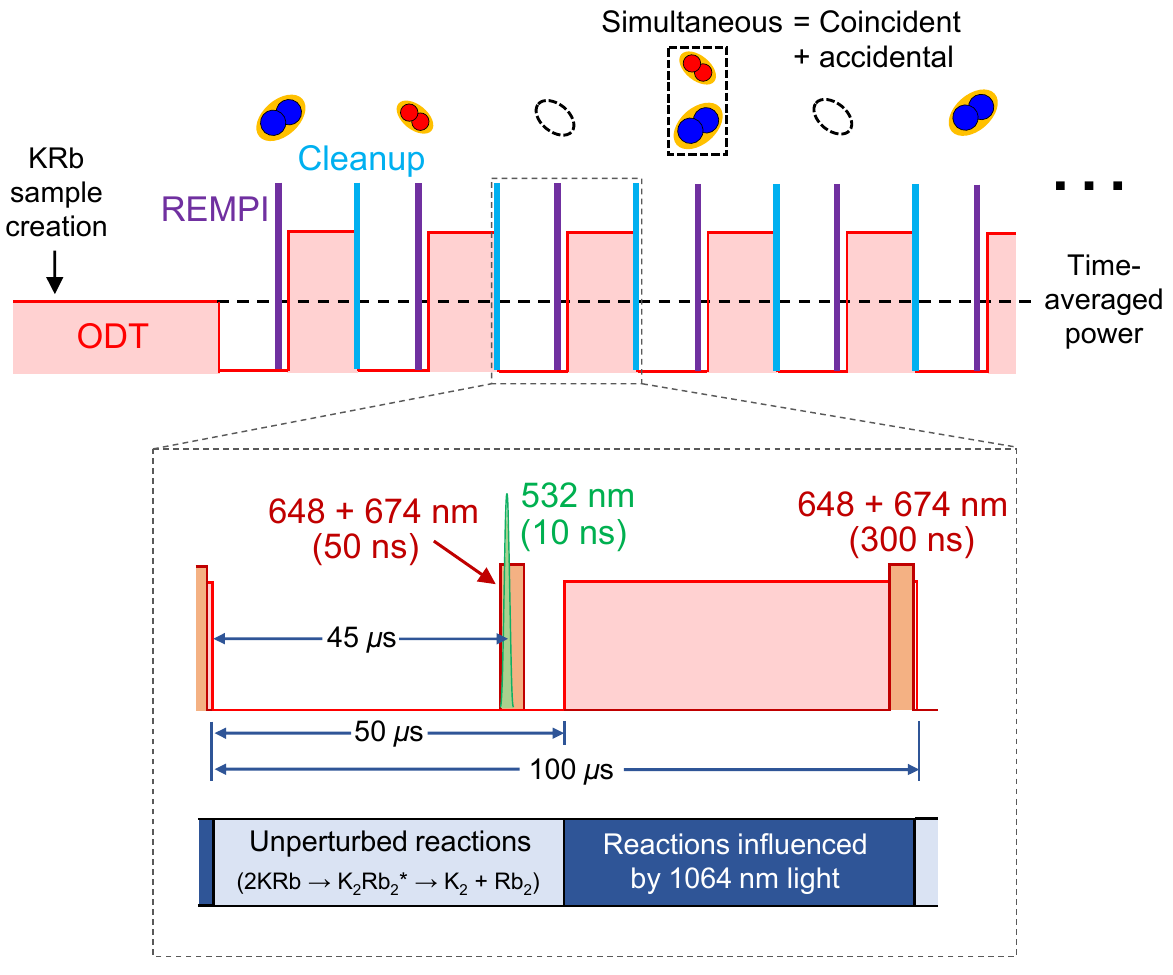}
    \caption{\normalsize\textbf{Timing diagram for product ionization}. The relative timing between the ODT, REMPI, and cleanup pulses during the state-selective ionization of reaction products. (Inset) A close up view of a single modulation period. Unperturbed reactions occur during the dark phase of the period, while reactions influenced by the (1064 nm) ODT light occur during the bright phase. The numbers in parentheses indicate pulse durations.}
\label{figIonTiming}
\end{figure}

Fig. \ref{figIonTiming} illustrates the timing of the various lasers involved in the state-selective ionization of the reaction products. The reactant KRb molecules are created inside an optical-dipole trap (ODT) that has a peak optical intensity of 11.3 kW/cm$^2$. Shortly after creation, a 50$\%$ duty-cycle square-wave modulation at a 10 kHz repetition rate is applied to the ODT intensity to create alternating bright and dark phases. This allows reactions to occur without the interference of the ODT light for half the time, while also maintaining a time-averaged trapping potential for the KRb sample. In previous work, it was shown that the ODT light strongly photo-excites the K$_2$Rb$_2^*$ complex, which influences the reaction pathway~\cite{liu2020photo}. During the dark phase of each ODT modulation period, 45 $\mu$s after the ODT turns off, we apply a resonance-enhanced multiphoton ionization (REMPI) pulse that consists of three wavelength components -- 648, 674, and 532 nm. When tuned to the appropriate rovibronic transition frequency, the 648 (674) nm light excites K$_2$ (Rb$_2$) from the selected rovibrational state $| \nu_{\textrm{K}_2} = 0, N_{\textrm{K}_2} \rangle$ ($| \nu_{\textrm{Rb}_2} = 0, N_{\textrm{Rb}_2} \rangle$). The 532 nm light then ionizes the excited molecules. Details of this ionization scheme are reported in Ref. ~\cite{hu2020product}.

During the bright phases of the ODT intensity modulation, reaction products are still being generated, albeit at a much lower rate compared to that during the dark phases.
The quantum states of these products are unknown, but can potentially include the target state-pair of our REMPI, $| N_{\textrm{K}_2}, N_{\textrm{Rb}_2} \rangle$. Since our goal is to ionize products of the unperturbed KRb + KRb reaction during the dark phases, the ionization of products in the same target state-pair produced during the bright phases will confound our measurement. To avoid this, we apply, at the end each bright phase, a ``cleanup'' pulse that consists of a pulse of 648 and 674 nm lights at the same frequencies used in the REMPI pulse. This will photo-excite products in $N_{\textrm{K}_2}$ and $N_{\textrm{Rb}_2}$, which, upon the ensuing spontaneous emission, will have negligible probabilities to decay back to the original states. As such, they will not be ionized by the following REMPI pulse. By monitoring the decay of the K$_2^+$ and Rb$_2^+$ signals over the course of the reaction, both in the presence and absence of the cleanup pulses, we found that these pulses do not noticeably perturb the reactants in the ODT.

\section{\large~Reaction exoergicity from literature} \label{section: exothermicity}

Knowledge of the reaction exoergicity, $\Delta E$, is important for our initial determination of the allowed rovibrational states of the products and their translational energies. To this end, we calculate it using known molecular dissociation energies, as
\begin{equation} \label{exoergicityEqn}
    \Delta E = D_0(\textrm{K}_2) + D_0(\textrm{Rb}_2) - 2D_0(\textrm{KRb}).
\end{equation}
Here, $D_0$ represents the dissociation energy, which is measured from the rovibrational ground-state of each species to the threshold of its dissociation into two free atoms. In the case of $^{40}$K$^{87}$Rb, Ni \textit{et al.}~\cite{ni2008high} obtained $D_0$ from the frequency difference between the lasers used to drive a molecular Raman transition, which were calibrated using a frequency comb to megahertz-level precision. In the cases of K$_2$ and Rb$_2$, large numbers of rovibronic transition frequencies were measured and used to fit the ground electronic potentials, yielding values for the well depth, $D_e$, that are accurate to better than 0.1 cm$^{-1}$. These fitted potentials allow for the calculation of the harmonic frequencies, $\omega_0$. The dissociation energies of K$_2$ and Rb$_2$ are then obtained by adding the zero-point-energy to the well depth, \textit{i.e.} $D_0 = D_e + \omega_0/2 $. The best known literature values for $D_e$, $\omega_0$, and $D_0$ are summarized in Tab. \ref{tab:D0}, along with their references.

\begin{table}
\centering
\begin{threeparttable}
\caption{\normalsize\label{tab:D0} Molecular dissociation energies of $^{40}$K$^{87}$Rb, $^{40}$K$_2$, and $^{87}$Rb$_2$. $D_e$: well-depth; $\omega_0$: vibrational constant; $D_0$: dissociation energy.}
\normalsize
\begin{tabular}{lcccc}
\hline\hline
Species & $D_e$ (cm$^{-1}$) & $\omega_0$ (cm$^{-1}$) & $D_0$ (cm$^{-1}$) & Reference \\
\hline
$^{40}$K$^{87}$Rb & & & -4180.442 & ~\cite{ni2008high} \\
$^{40}$K$_2$ & -4450.904(4) & 91.032(1) & -4405.389(4) & ~\cite{falke2006sigma} \\
$^{87}$Rb$_2$ & -3993.53(6) & 57.121(1) & -3964.97(6) & ~\cite{seto2000direct} \\
\hline\hline
\end{tabular}
\end{threeparttable}
\end{table}

Using the tabulated values for $D_0$ and Eq. \ref{exoergicityEqn}, we calculate an exoergicity of $\Delta E = -9.53(7) $ cm$^{-1}$.
We note that this is lower in absolute value compared to the assumed $- 10.4(4)$ cm$^{-1}$ calculated in Ref. ~\cite{ospelkaus2010quantum}, which used an earlier literature value for $D_0(\textrm{Rb}_2)$~\cite{amiot1990laser}. As such, we use $\Delta E \sim -10 $ cm$^{-1}$ as a rough initial estimate.

\section{\large~Internal energy of a state-pair} \label{section: internal energy}

The internal energy of a product pair $| N_{\textrm{K}_2}, N_{\textrm{Rb}_2} \rangle $ in the vibrational ground state ($| \nu_{\textrm{K}_2} = 0, \nu_{\textrm{Rb}_2} = 0 \rangle $) is given by
\begin{equation} \label{EnergyCon}
\begin{split}
    U(N_{\textrm{K}_2}, N_{\textrm{Rb}_2}) &= B_{\textrm{K}_2}N_{\textrm{K}_2}(N_{\textrm{K}_2}+1) - D_{\textrm{K}_2}\left(N_{\textrm{K}_2}(N_{\textrm{K}_2}+1)\right)^2 \\
    & +  B_{\textrm{Rb}_2}N_{\textrm{Rb}_2}(N_{\textrm{Rb}_2}+1) - D_{\textrm{Rb}_2}\left(N_{\textrm{Rb}_2}(N_{\textrm{Rb}_2}+1)\right)^2,
\end{split}
\end{equation}
Here, $B$ and $D$ are the rotational and centrifugal constants, respectively. The best known literature values for $B$ and $D$ are listed in Tab. \ref{tab: rovibrational constants}, along with their references. Note that they are scaled by mass from the values measured for their more abundant isotopic counterparts, according to $B_b = B_a (\mu_a/\mu_b)$ and $D_b = D_a (\mu_a/\mu_b)^2$. Here, $\mu$ represents the reduced mass, $a$ represents $^{85}$Rb$_2$ or $^{39}$K$_2$, $b$ represents $^{87}$Rb$_2$ or $^{40}$K$_2$. 
The internal energies for all state-pairs relevant to this study are listed in Tab \ref{tab:productStates_theory}.

\begin{table}[t]
\centering
\begin{threeparttable}
\caption{\normalsize\label{tab: rovibrational constants} Rotational and centrifugal constants for $^{40}$K$_2$ and $^{87}$Rb$_2$, scaled by mass from the measured values for $^{39}$K$_2$ and $^{85}$Rb$_2$.}
\normalsize
\begin{tabular}{lccc}
\hline\hline
Quantity & Symbol & Value (cm$^{-1}$) & Reference \\
\hline
$^{40}$K$_2$ rotational constant & $B_{\textrm{K}_2}$ & $5.478155(84)\times10^{-2}$ & ~\cite{amiot1995long} \\
$^{40}$K$_2$ centrifugal constant & $D_{\textrm{K}_2}$ & $7.8641(68)\times10^{-8}$ & ~\cite{amiot1995long} \\
$^{87}$Rb$_2$ rotational constant & $B_{\textrm{Rb}_2}$ & $2.188943(61)\times10^{-2}$ & ~\cite{seto2000direct} \\
$^{87}$Rb$_2$ centrifugal constant & $D_{\textrm{Rb}_2}$ & $1.29507(56)\times10^{-8}$ & ~\cite{seto2000direct} \\
\hline\hline
\end{tabular}
\end{threeparttable}
\end{table}

\section{\large~Counting the number of scattering channels within a state-pair} \label{section: number of channels}

For a given state-pair $| N_{\textrm{K}_2}, N_{\textrm{Rb}_2} \rangle$, additional scattering channels arise due to the freedom in choosing the relative orientations of the corresponding rotation vectors $\vec{N}_{\textrm{K}_2}$ and $\vec{N}_{\textrm{Rb}_2}$. The number of such channels can be determined with the assumption of total angular momentum conservation throughout the reaction,
\begin{equation} \label{AngMomCon}
    \vec{J}_{\textrm{tot}} = \vec{N}_{\textrm{reac}} + \vec{L}_{\textrm{reac}} = \vec{N}_{\textrm{prod}} + \vec{L}_{\textrm{prod}},
\end{equation}
where
\begin{equation} \label{CoupledAngMom}
    \begin{aligned}
        \vec{N}_{\textrm{reac}} &= \vec{N}_{\textrm{KRb(1)}} + \vec{N}_{\textrm{KRb(2)}} \\
        \vec{N}_{\textrm{prod}} &= \vec{N}_{\textrm{K}_2} + \vec{N}_{\textrm{Rb}_2}.
\end{aligned}
\end{equation}
Here, $\vec{J}_{\textrm{tot}}$ represents the total angular momentum of the system, $\vec{N}_{\textrm{reac}}$ ($\vec{N}_{\textrm{prod}}$) represents the coupled rotational angular momenta of the reactants (products), and $\vec{L}_{\textrm{reac}}$ ($\vec{L}_{\textrm{prod}}$) represents the orbital angular momenta of the reactants (products).
The above relations between angular momentum vectors are equivalent to a set of triangle inequalities on the corresponding quantum numbers, written as
\begin{equation} \label{triIneq}
    \begin{aligned}
         |N_{\textrm{KRb(1)}} - N_{\textrm{KRb(2)}}| \leq &N_{\textrm{reac}} \leq |N_{\textrm{KRb(1)}} + N_{\textrm{KRb(2)}}|, \\
         |N_{\textrm{K}_2} - N_{\textrm{Rb}_2}| \leq &N_{\textrm{prod}} \leq |N_{\textrm{K}_2} + N_{\textrm{Rb}_2}|, \\
         |N_{\textrm{reac}} - L_{\textrm{reac}}| \leq &J_{\textrm{tot}} \leq |N_{\textrm{reac}} + L_{\textrm{reac}}|, \\
        |L_{\textrm{prod}} - N_{\textrm{prod}}| \leq &J_{\textrm{tot}} \leq |L_{\textrm{prod}} + N_{\textrm{prod}}|.
    \end{aligned}
\end{equation}

Due to the presence of a 30 G magnetic field during the reaction (Fig. \ref{figDetectionScheme}A), an additional constraint on the product angular momentum quantum numbers is imposed by the conservation of total parity~\cite{gonzalez2014statistical},
\begin{equation} \label{ParityCon}
    (-1)^{N_{\textrm{KRb(1)}}}(-1)^{N_{\textrm{KRb(2)}}}(-1)^{L_{\textrm{reac}}} = (-1)^{N_{\textrm{K}_2}}(-1)^{N_{\textrm{Rb}_2}}(-1)^{L_{\textrm{prod}}}.
\end{equation}

\begin{table}[t!]
\centering
\caption{\normalsize\normalsize\label{tab:productStates_theory} The internal energy ($U$, section \ref{section: internal energy}) and degeneracy ($\mathcal{D}$, section \ref{section: number of channels}) of all measured product state-pairs.}
\normalsize
\begin{threeparttable}
\begin{tabular}{llll|llll|llll}
\hline
$N_{\textrm{K}_2}$ & $N_{\textrm{Rb}_2}$ & $U$(cm$^{-1}$) & $\mathcal{D}$ & $N_{\textrm{K}_2}$ & $N_{\textrm{Rb}_2}$ & $U$(cm$^{-1}$) & $\mathcal{D}$ & $N_{\textrm{K}_2}$ & $N_{\textrm{Rb}_2}$ & $U$(cm$^{-1}$) & $\mathcal{D}$\\
\hline\hline
0 & 1 & 0.043779 & 2 & 4 & 1 & 1.1394 & 5 & 8 & 3 & 4.2065 & 11 \\
0 & 3 & 0.26267 & 2 & 4 & 3 & 1.3583 & 11 & 8 & 5 & 4.6005 & 17 \\
0 & 5 & 0.65667 & 2 & 4 & 5 & 1.7523 & 14 & 8 & 7 & 5.1696 & 23 \\
0 & 7 & 1.2258 & 2 & 4 & 7 & 2.3214 & 14 & 8 & 9 & 5.9138 & 26 \\
0 & 9 & 1.9699 & 2 & 4 & 9 & 3.0655 & 14 & 8 & 11 & 6.8330 & 26 \\
0 & 11 & 2.8892 & 2 & 4 & 11 & 3.9848 & 14 & 8 & 13 & 7.9273 & 26 \\
0 & 13 & 3.9834 & 2 & 4 & 13 & 5.0790 & 14 & 8 & 15 & 9.1966 & 26 \\
0 & 15 & 5.2527 & 2 & 4 & 15 & 6.3483 & 14 & 10 & 1 & 6.0688 & 5 \\
0 & 17 & 6.6970 & 2 & 4 & 17 & 7.7926 & 14 & 10 & 3 & 6.2877 & 11 \\
0 & 19 & 8.3161 & 2 & 4 & 19 & 9.4117 & 14 & 10 & 5 & 6.6817 & 17 \\
2 & 1 & 0.3725 & 5 & 6 & 1 & 2.3445 & 5 & 10 & 7 & 7.2508 & 23 \\
2 & 3 & 0.5914 & 8 & 6 & 3 & 2.5634 & 11 & 10 & 9 & 7.9950 & 29 \\
2 & 5 & 0.9854 & 8 & 6 & 5 & 2.9574 & 17 & 10 & 11 & 8.9142 & 32 \\
2 & 7 & 1.5545 & 8 & 6 & 7 & 3.5265 & 20 & 10 & 13$^\dag$ & 10.0085 & 32 \\
2 & 9 & 2.2986 & 8 & 6 & 9 & 4.2706 & 20 & 10 & 15$^\dag$ & 11.2777 & 32 \\
2 & 11 & 3.2179 & 8 & 6 & 11 & 5.1899 & 20 & 12 & 1 & 8.5878 & 5 \\
2 & 13 & 4.3121 & 8 & 6 & 13 & 6.2841 & 20 & 12 & 3 & 8.8067 & 11 \\
2 & 15 & 5.5814 & 8 & 6 & 15 & 7.5534 & 20 & 12 & 5 & 9.2007 & 17 \\
2 & 17 & 7.0256 & 8 & 6 & 17 & 8.9976 & 20 & 12 & 7 & 9.7698 & 23 \\
2 & 19 & 8.6448 & 8 & 8 & 1 & 3.9876 & 5 &  &  &  & \\
\hline
\end{tabular}
$^\dag$ Energetically forbidden \\
\end{threeparttable}
\end{table}

Because the KRb reactants in our experiments are prepared in their rovibrational ground state and collide via $p$-wave collisions, we have $N_{\textrm{KRb(1)}} = N_{\textrm{KRb(2)}} = 0$, and therefore $J_{\textrm{tot}} = L_{\textrm{reac}} = 1$. Given this initial condition, we count, for each given pair of $| N_{\textrm{K}_2}, N_{\textrm{Rb}_2}\rangle$, the number of $| N_{\textrm{prod}}, L_{\textrm{prod}}\rangle$ combinations that satisfy Eq. \ref{triIneq} and \ref{ParityCon} to obtain its degeneracy ($\mathcal{D}$). The results are documented in Tab. \ref{tab:productStates_theory}. As an example, for the state, $|N_{\textrm{K}_2} = 2, N_{\textrm{Rb}_2} = 1\rangle$, there exists 5 channels, which are $|N_{\textrm{prod}}, L_{\textrm{prod}}\rangle$ = $|1, 0\rangle,|1, 2\rangle,|2, 2\rangle,|3, 2\rangle$, and $|3, 4\rangle$. Note that the nuclear spins and their associated angular momenta are ignored for the purpose of this state-counting. This is justified by the results of our previous work, in which it was shown that the nuclear spins remain unchanged throughout the reaction, and are therefore effectively decoupled from the dynamics~\cite{hu2020product}.

The state counting arguments given above rely on the assumption that the different combinations of $| N_{\textrm{prod}}, L_{\textrm{prod}}\rangle$ which satisfy Eq. \ref{triIneq} and \ref{ParityCon} for a given pair of $| N_{\textrm{K}_2}, N_{\textrm{Rb}_2}\rangle$ are effectively degenerate in energy. In other words, the hyperfine structures of the K$_2$ and Rb$_2$ product molecules, along with any corresponding energy splittings that would break the degeneracy of the different $| N_{\textrm{prod}}, L_{\textrm{prod}}\rangle$ combinations associated with each $| N_{\textrm{K}_2}, N_{\textrm{Rb}_2}\rangle$ pair, have been ignored. For the purposes of this work, this assumption is justified as the energy splittings associated with the hyperfine structure of each product species are significantly smaller than the spectral resolution of the REMPI detection used in the experiment, which is approximately $45$ MHz ~\cite{hu2020product}. 

Specifically, using the hyperfine Hamiltonian and calculated hyperfine coupling constants reported in Ref.~\cite{Aldegunde2009}, along with the rotational and centrifugal constants given in Table \ref{tab: rovibrational constants}, we have calculated the spectral width of each rotational manifold. This is accomplished by diagonalizing the Hamiltonian in the presence of a $30$ G magnetic field for each product species and including rotational states up to $N_{\textrm{K}_2}=12$ for the K$_2$ products and $N_{\textrm{Rb}_2}=19$ for the Rb$_2$ products. The spectral width here is defined as the energy difference between the highest energy hyperfine state and the lowest energy hyperfine state associated with a particular rotational quantum number. From the results of these calculations, we find that the spectral widths of the product rotational states that are relevant to this work, which arise from the hyperfine structure of each product species, are less than $1.11$ MHz for Rb$_2$ and less than $0.27$ MHz for K$_2$. Because this is significantly smaller than the experimental resolution of $45$ MHz, the hyperfine structure of the products is not resolved and the different combinations of $| N_{\textrm{prod}}, L_{\textrm{prod}}\rangle$ associated with each pair of $| N_{\textrm{K}_2}, N_{\textrm{Rb}_2}\rangle$ can be considered degenerate.

We also note here that the spectral width associated with the hyperfine structure of the product molecules is smaller than the experimental uncertainty in the reaction exoergicity, $9.7711^{+0.0007}_{-0.0005}$ cm$^{-1}$, reported in the main text. In units of frequency, this uncertainty corresponds to an upper error bound of $20$ MHz and a lower bound of $15$ MHz, which are both larger than the hyperfine width of the product rotational states. We have therefore ignored the hyperfine structure of the products in determining the reaction exoergicity in the main text.

\section{\large~State-selective coincidence detection of product pairs} \label{section: mtm filtering}

Coincidence imaging is a powerful tool for simultaneously probing multiple product molecules from individual reaction events. The driving force behind this technique is our three-dimensional detection system which is capable of measuring both the TOF and the transverse velocity of product ions. This information can be used to extract the three-dimensional momentum vectors of the initial product molecules, which enables the identification of coincident K$_2$ and Rb$_2$ product pairs based on the correlations of their momenta~~\cite{lee2014coincidence,ullrich2003recoil,vredenborg2008photoelectron}. In this section, we describe a method to perform state-resolved coincidence imaging of product molecules which combines REMPI spectroscopy with velocity map imaging (VMI) of ions.
The former technique provides the capability to resolve the different product rotational states by taking advantage of the unique resonance frequencies of the corresponding bound-to-bound molecular transitions. The latter enables the detection of coincident product pairs, and therefore the resolution of individual product state-pairs, by providing access to the momentum information of the product molecules.

VMI ion optics map the momenta of photo-ionized neutrals onto spatial locations that can be measured by a position-sensitive ion detector.
The configuration of the VMI optics consists of three main electrode plates (Fig. \ref{figDetectionScheme}A): a repeller plate, an extractor plate, and a ground plate, with the geometries and voltages of these electrodes chosen to optimize the performance of the imaging system~~\cite{eppink1997velocity}. The ion detector used in our experiment is a delay-line MCP (Roentdek DLD80), which has an active diameter of 80 mm, a spatial resolution of 0.08 mm, and a temporal resolution of 1 ns. Additional details of the ionization and detection setup are reported in Ref.~~\cite{liu2020probing}.
Before the ionization process, a small volume of neutral product molecules resides in the area between the repeller and extractor electrodes. Upon the ionization of these molecules, the charged species are accelerated towards the ion detector along the TOF axis, while they simultaneously expand ballistically at a rate which is determined by the initial transverse velocities of the corresponding neutral molecules.
This results in a mapping of the transverse momentum of each neutral product ($p_x$, $p_y$) onto its impact position on the detector ($X$, $Y$) according to the relations $p_{\textrm{s},x} \propto \sqrt{2 m_\textrm{s}} (X_\textrm{s} - X_{\textrm{s}}^0) = \sqrt{2 m_\textrm{s}} \Delta X_s $ and $p_{\textrm{s},y} \propto \sqrt{2 m_\textrm{s}} (Y_\textrm{s} -  Y_{\textrm{s}}^0) = \sqrt{2 m_\textrm{s}} \Delta Y_s $. Here, $\textrm{s}$ represents the product species (K$_2$ or Rb$_2$), $m$ represents molecular mass, and $\{X_\textrm{s}^0,Y_\textrm{s}^0\}$ represents the impact position of zero-velocity products that is in general shifted from the detector center due to the presence of Lorentz forces during the ion flight. The axial momentum of a product ($p_z$), on the other hand, is mapped into its TOF as a result of its location relative to the center between the repeller and extractor plates ~\cite{hu2019direct}, according to the relation $p_{\textrm{s},z} \propto (TOF_{\textrm{s}} - TOF_{\textrm{s}}^0)/(\eta\Delta t/TOF_{\textrm{s}}^0-1) = \Delta TOF_{\textrm{s}} /(\eta\Delta t/TOF_{\textrm{s}}^0-1)$. Here, $TOF^0$ represents the TOF for products with zero initial velocity, which is $69.27$ and $102.13$ $\mu$s for K$_2$ and Rb$_2$, respectively; $\Delta t$ represents the time between the initial formation of the K$_2$ and Rb$_2$ product pair and its subsequent ionization (Fig. \ref{figIonTiming}), which can take any value in the range $0 - 45~\mu$s; and $\eta$ is a dimensionless parameter determined by the geometry of our electrodes, whose value is 136.

\begin{table}[p]
\centering
\caption{\normalsize\normalsize \label{tab:productStates_exp} The experimental cycle number ($C_{\textrm{exp}}$), simultaneous counts ($n_{\textrm{sim}}$), background level ($n_{\textrm{bkgd}}$), and background fluctuation ($\delta n_{\textrm{bkgd}}$) for with each measured product state-pair. The values for $n_{\textrm{sim}}$, $n_{\textrm{bkgd}}$, and $\delta n_{\textrm{bkgd}}$ are obtained after the momentum-based screening is applied (section S5). The coincidence counts for each state-pair is calculated as $n_{\textrm{coin}} = n_{\textrm{sim}} - n_{\textrm{bkgd}}$. Also included are the (base 10) logarithm of the state-specific $p$-values used towards hypothesis testing (section \ref{likelihood ratio test}).}
\normalsize
\begin{threeparttable}
\begin{tabular}{lllllll|lllllll}
\hline
$N_{\textrm{K}_2}$ & $N_{\textrm{Rb}_2}$ & $C_{\textrm{exp}}$ & $n_{\textrm{sim}}$ & $n_{\textrm{bkgd}}$ & $\delta n_{\textrm{bkgd}}$ & $\log(p)$ & $N_{\textrm{K}_2}$ & $N_{\textrm{Rb}_2}$ & $C_{\textrm{exp}}$ & $n_{\textrm{sim}}$ & $n_{\textrm{bkgd}}$ & $\delta n_{\textrm{bkgd}}$ & $\log(p)$ \\
\hline\hline
0 & 1 & 1947 & 2 & 0.2 & 0.4 & -2.8 & 6 & 1 & 1629 & 17 & 0.6 & 0.7 & -0.3 \\
0 & 3 & 1603 & 11 & 0.6 & 0.7 & -0.2 & 6 & 3 & 820 & 18 & 1 & 1 & 0.0 \\
0 & 5 & 1402 & 2 & 0.6 & 0.7 & -1.3 & 6 & 5 & 859 & 34 & 2.8 & 1.7 & -0.1 \\
0 & 7 & 1871 & 11 & 0.5 & 0.8 & -0.6 & 6 & 7 & 1372 & 91 & 3.4 & 1.7 & 0.0 \\
0 & 9 & 1984 & 16 & 1.1 & 1 & -1.0 & 6 & 9 & 989 & 73 & 4.1 & 1.8 & -0.3 \\
0 & 11 & 1998 & 24 & 1.4 & 1.3 & -2.4 & 6 & 11 & 957 & 86 & 7 & 2.7 & -0.4 \\
0 & 13 & 2614 & 31 & 1.2 & 1.1 & -1.5 & 6 & 13 & 935 & 56 & 3.3 & 1.7 & -2.5 \\
0 & 15 & 2019 & 29 & 2 & 1.4 & -0.5 & 6 & 15 & 968 & 220 & 8.7 & 2.7 & -1.6 \\
0 & 17 & 998 & 12 & 0.4 & 0.7 & -0.5 & 6 & 17 & 638 & 123 & 3.3 & 1.9 & -1.7 \\
0 & 19 & 1000 & 21 & 0.2 & 0.4 & -0.9 & 8 & 1 & 1594 & 30 & 2.4 & 1.6 & -0.1 \\
2 & 1 & 1551 & 5 & 0.3 & 0.6 & -2.1 & 8 & 3 & 1021 & 52 & 4.2 & 2.2 & -0.1 \\
2 & 3 & 918 & 12 & 0.5 & 0.8 & -0.1 & 8 & 5 & 1011 & 74 & 10.1 & 3.8 & -0.4 \\
2 & 5 & 1016 & 21 & 1.7 & 1.2 & -0.3 & 8 & 7 & 1018 & 137 & 6.1 & 2.2 & -0.5 \\
2 & 7 & 1067 & 13 & 1.4 & 1.3 & -1.4 & 8 & 9 & 1260 & 228 & 10.5 & 3 & -1.5 \\
2 & 9 & 994 & 26 & 1.9 & 1.5 & 0.0 & 8 & 11 & 962 & 145 & 10.5 & 3.1 & -0.2 \\
2 & 11 & 1051 & 25 & 3.1 & 1.7 & -0.4 & 8 & 13 & 1050 & 254 & 7.2 & 2.2 & -0.6 \\
2 & 13 & 1398 & 53 & 2.6 & 1.6 & -0.2 & 8 & 15 & 1334 & 684 & 17.6 & 3.9 & -11 \\
2 & 15 & 1028 & 39 & 3 & 1.6 & -0.5 & 10 & 1 & 2006 & 33 & 2.8 & 1.7 & -4.1 \\
2 & 17 & 1127 & 47 & 2 & 1.2 & -0.4 & 10 & 3 & 980 & 26 & 2.7 & 1.6 & -2.3 \\
2 & 19 & 1614 & 144 & 2 & 1.3 & -0.6 & 10 & 5 & 1051 & 122 & 8.9 & 3 & -0.1 \\
4 & 1 & 1046 & 7 & 0.4 & 0.8 & -0.4 & 10 & 7 & 975 & 151 & 4.8 & 2.5 & -0.5 \\
4 & 3 & 1697 & 28 & 2.3 & 1.5 & -0.6 & 10 & 9 & 1048 & 287 & 9.7 & 3 & 0.0 \\
4 & 5 & 1021 & 43 & 3.8 & 2 & -0.4 & 10 & 11 & 982 & 428 & 11.1 & 3.3 & -0.3\\
4 & 7 & 1416 & 42 & 3.2 & 1.6 & -0.7 & 10 & 13$^{\dag}$ & 1039 & 10 & 7.7 & 2.9 &  \\
4 & 9 & 979 & 44 & 3.2 & 1.7 & -0.4 & 10 & 15$^{\dag}$ & 854 & 9 & 10.9 & 3.2 &  \\
4 & 11 & 1456 & 112 & 8.8 & 2.8 & -1.1 & 12 & 1 & 1988 & 106 & 1.2 & 1 & -0.5 \\
4 & 13 & 962 & 67 & 4.4 & 2 & -0.3 & 12 & 3 & 904 & 119 & 2 & 1.4 & -1.4 \\
4 & 15 & 1063 & 101 & 8.2 & 2.8 & -0.1 & 12 & 5 & 1032 & 311 & 4 & 2 & -6.1 \\
4 & 17 & 1062 & 112 & 3.6 & 1.9 & -0.3 & 12 & 7 & 2544 & 31 & 5.3 & 2.1 & -18 \\
4 & 19 & 909 & 207 & 2.6 & 2 & -2.4 &  &  &  &  &  &  & \\
\hline
\end{tabular}
$^\dag$ Energetically forbidden \\
\end{threeparttable}
\end{table}

We identify coincidence ion pairs as those that satisfy the momentum conservation condition $\vec{p}_{\text{K}_2}+\vec{p}_{\text{Rb}_2} = 0$, or equivalently $p_{\text{K}_2,x}+p_{\text{Rb}_2,x} = 0$, $p_{\text{K}_2,y}+p_{\text{Rb}_2,y} = 0$, and $p_{\text{K}_2,z}+p_{\text{Rb}_2,z} = 0$. Given how an product molecule's momenta are related to its TOF and impact position, as well as the finite position and timing resolutions of our ion imaging system, the conservation of momentum translates into a set of screening criteria,
\begin{eqnarray}
\left|\Delta X_{\text{K}_2}+\Delta X_{\text{Rb}_2}\sqrt{m_{\text{Rb}_2}/m_{\text{K}_2}} \right|&\leq& n\sigma_X\sqrt{1+m_{\text{Rb}_2}/m_{\text{K}_2}}, \label{eq-X}\\
\left|\Delta Y_{\text{K}_2}+\Delta Y_{\text{Rb}_2}\sqrt{m_{\text{Rb}_2}/m_{\text{K}_2}} \right|&\leq& n\sigma_Y\sqrt{1+m_{\text{Rb}_2}/m_{\text{K}_2}},  \label{eq-Y} \\
\left|\Delta TOF_{\text{K}_2}+\Delta TOF_{\text{Rb}_2}\frac{\eta\Delta t/TOF^0_{\text{K}_2}-1}{\eta\Delta t/TOF^0_{\text{Rb}_2}-1} \right|&\leq& n\sigma_T\sqrt{1+\left(\frac{\eta\Delta t/TOF^0_{\text{K}_2}-1}{\eta\Delta t/TOF^0_{\text{Rb}_2}-1} \right)^2}.  \label{eq-TOF}
\end{eqnarray}
Here, $\sigma_{X,Y,T}$ represent the $1\sigma$ resolution of our detection system along the $X$, $Y$, and $TOF$ axes, respectively, which are measured to be 0.23 mm, 0.23 mm and 11 ns. The multiplication factor $n$ is empirically determined to be 3.

The efficacy of the above screening process is manifested in the process's ability to discriminate against uncorrelated ion counts, which we demonstrate in Fig. \ref{figDetectionScheme} using the product ion data for $| N_{\text{K}_2}=6, N_{\text{Rb}_2}=7 \rangle$ as an example. Fig. \ref{figDetectionScheme}D shows the total number of detected K$_2^+$-Rb$_2^+$ pairs generated by two ionization pulses which are separated from one another by $k$ pulses. There, one observes a prominent peak at $k = 0$, which corresponds to the number of simultaneously detected ion pairs, $n_{\textrm{sim}}$. The measured counts with $k \neq 0$, which correspond to uncorrelated ion pairs generated by separate pulses, form a uniform background with a mean value of $n_{\textrm{bkgd}}$ and a standard deviation of $\delta n_{\textrm{bkgd}}$. Fig. \ref{figDetectionScheme}E shows the counts remaining after all ion pairs are subjected to screening based on Eqs. \ref{eq-X}-\ref{eq-TOF}. There, one observes a strong suppression of the background level, indicating that the uncorrelated pairs are effectively screened away. The height of the $k = 0$ peak is reduced, as accidental counts are screened away, while true coincidence counts remain. Since background level remains finite after screening, we use it as a proxy for the number of accidental counts that remain, and subtract it off to obtain the true coincidence counts, $n_{\textrm{coin}} = n_{\textrm{sim}} - n_{\textrm{bkgd}}$.
The measured values of $n_{\textrm{sim}}$, $n_{\textrm{bkgd}}$, and $\delta n_{\textrm{bkgd}}$ for all state-pairs are listed in Tab. \ref{tab:productStates_exp}.

The uncertainty for the coincidence counts, $\delta n_{\textrm{coin}}$, has contributions from three sources -- the shot noise associated with the simultaneous ion counts ($\sqrt{n_{\textrm{sim}}}$), the fluctuation of the background level ($\delta n_{\textrm{bkgd}}$), and the fluctuation in experimental conditions ($\alpha \times n_{\textrm{coin}}$, see section \ref{section: normalization of coincidence counts}B). Since these errors are uncorrelated, they are summed in quadrature to yield the overall uncertainty, $\delta n_{\textrm{coin}} = \sqrt{n_{\textrm{sim}} + (\delta n_{\textrm{bkgd}})^2 + (\alpha \times n_{\textrm{coin}})^2}$. For state-pairs with sufficient statistics ($n_{\textrm{sim}} \geq 5 $), which makes up the vast majority of the measured state-pairs, the contribution of $\delta n_{\textrm{bkgd}}$ to the overall uncertainty is very small ($< 5\%$). The values for $\delta n_{\textrm{coin}}$ are reflected by the error bars in Fig. \ref{figProdStateDist}B-H, whose sizes are $\pm\delta n_{\textrm{coin}}$. This uncertainty is propagated into that of the normalized coincidence counts as well as the measured scattering probabilities that constitute the product state distribution ($P_{\textrm{sc}}^{\textrm{meas}}$).

We note that while methods for obtaining coincident quantum state information for a pair of products already exist~~\cite{gericke1988correlations,lin2003state}, their resolution is insufficient for resolving the small spacing between rotational levels of heavy molecules such as K$_2$ and Rb$_2$. 
Hence the current scheme represents a new approach to complete product state detection that is generally applicable to reactions involving polyatomic species.

\section{\large~Normalization of coincidence counts} \label{section: normalization of coincidence counts}

The screening process described in section \ref{section: mtm filtering} allows us to extract the number of coincidence ion pair counts, $n_{\textrm{coin}}(N_{\textrm{K}_2}, N_{\textrm{Rb}_2})$, from the data set associated with each given state-pair. In order for $n_{\textrm{coin}}(N_{\textrm{K}_2}, N_{\textrm{Rb}_2})$ to proportionally reflect the scattering probability into $| N_{\textrm{K}_2}, N_{\textrm{Rb}_2} \rangle$, however, it must undergo normalizations against experimental biases that differ from one data set to another. Sources for these biases include the number of experimental cycles associated with each data set, fluctuations in experimental conditions between data sets, and the product-velocity-dependent efficiency of our REMPI sampling. In this section, we describe the procedures used to account for these biases.

\subsection{\large Normalization against product-velocity-dependent ionization sampling efficiency}

In our experiment, products generated by the reaction are sampled in a state-dependent fashion using REMPI. The efficiency of this sampling depends on the product velocities due to two mechanisms: 1. high velocity products have a higher chance of escaping the volume covered by the REMPI beams before the lights are pulsed on; and 2. products with velocity components along or against the direction of REMPI beam propagation will experience Doppler shift to the bound-to-bound transition frequency, which affects the probability for the product to be promoted to the excited state. We can respectively quantify the degree to which these two mechanisms affect the sampling of correlated product pairs using a geometric factor, $F_{\textrm{geometry}}(T)$, and a Doppler factor, $F_{\textrm{Doppler}}(T)$. Here, $T$ represents the translational energy of products in a given state-pair, and is related to the product velocities as $v_{\textrm{K}_2}(T) = \sqrt{2 T m_{\textrm{Rb}_2}/(m_{\textrm{Rb}_2} + m_{\textrm{K}_2})}$ and $v_{\textrm{Rb}_2}(T) = \sqrt{2 T m_{\textrm{K}_2}/(m_{\textrm{Rb}_2} + m_{\textrm{K}_2})}$.
In this section, we develop models for these factors.

\begin{figure}[t!]
\centering
\includegraphics[width=0.75\textwidth]{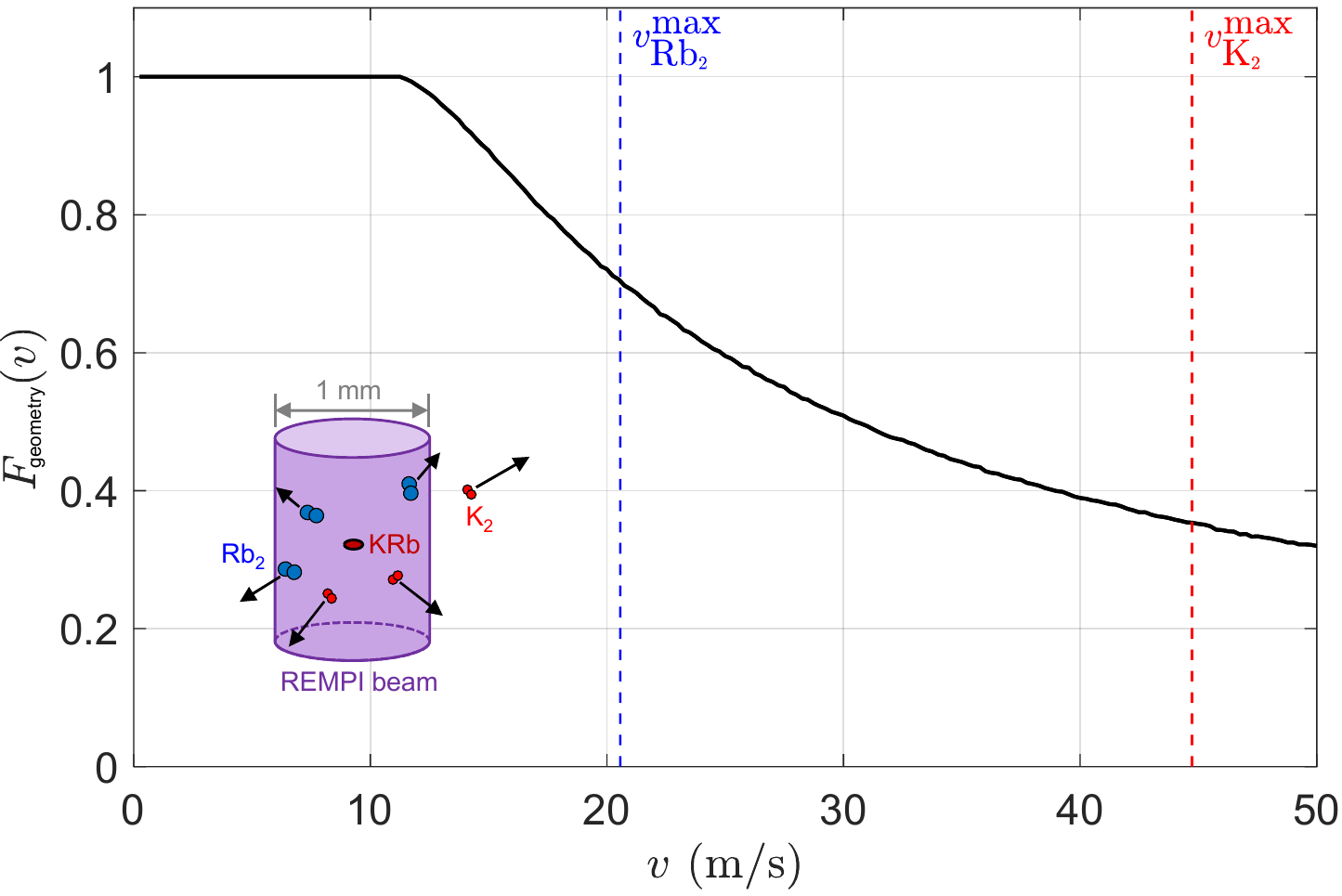}
    \caption{\normalsize\textbf{Modeling the geometric factor for product sampling}. The plot shows the fraction of product pairs that remain within the REMPI beam volume at the time of ionization as a function of the velocity of the K$_2$ product within the pair. Blue and red dashed lines indicate the maximum velocities of the K$_2$ and Rb$_2$ products, respectively. (inset) the ionization geometry.}
\label{figGeometricFactor}
\end{figure}

Physically, the geometric factor represent the fraction of product pairs that are generated during the dark phase of the ODT modulation (Fig. \ref{figIonTiming}), and remain inside the volume exposed to the REMPI beam when it is pulsed on. Over the region of interest, our REMPI beam can be approximated as a 1mm-diameter cylinder with a uniform intensity (for each frequency component). Since a K$_2$ product molecule will always travel faster compared to its Rb$_2$ coproduct, $F_{\textrm{geometry}}$ is determined solely by the fraction of K$_2$ molecules that remain within the beams given their velocity. To model this factor, we developed a numerical simulation that generates and samples reaction products according to the beam geometry and timing diagram shown in the insets of Fig. \ref{figGeometricFactor}.
The simulation takes the velocity of the products and other parameters from the experiment (\textit{i.e.} repetition rate $f$ and details of the timing) as inputs, and reports the fraction of products exposed to the REMPI beam over many detection periods. In brief, each period of the REMPI detection begins with the turn off of the ODT confining the KRb molecules, at which point products begin to emerge from reactions with velocity $v$ and propagate outward; after a time $t_{\textrm{del}}$, the REMPI beams are pulsed on, and the number of products under its exposure is accumulated; this period is repeated several hundred times to collect statistics. In the end, the number of exposed products is divided by the total number of products generated to obtain the exposed fraction $F_{\textrm{geometry}}(v)$. 

Fig. \ref{figGeometricFactor} shows $F_{\textrm{geometry}}(v)$ over the range of expected velocities for K$_2$ ($v = $ 0 - 44 m/s). It is calculated according to the timing scheme used in the experiment (see Fig. \ref{figIonTiming}), \textit{i.e.} $f = 10$ kHz, $t_{\textrm{del}} = 45 \mu$s, and a $50 \%$ duty-cycle for the ODT modulation. Products with $v < 11.1 $ m/s are fully contained within the cylindrical volume before ionization, and therefore have $F_{\textrm{geo}}(v) = 1$; those with $v > 11.1 $ m/s, on the other hand, experience a decay in $F_{\textrm{geo}}(v)$ that approximately scales as $1/v$. Using the simulation result, we calculated, for all allowed state-pairs, the geometric factor relevant for the normalization of coincidence counts ($F_{\textrm{geometry}}(T))$.

\begin{figure}[t!]
\centering
\includegraphics[width=1\textwidth]{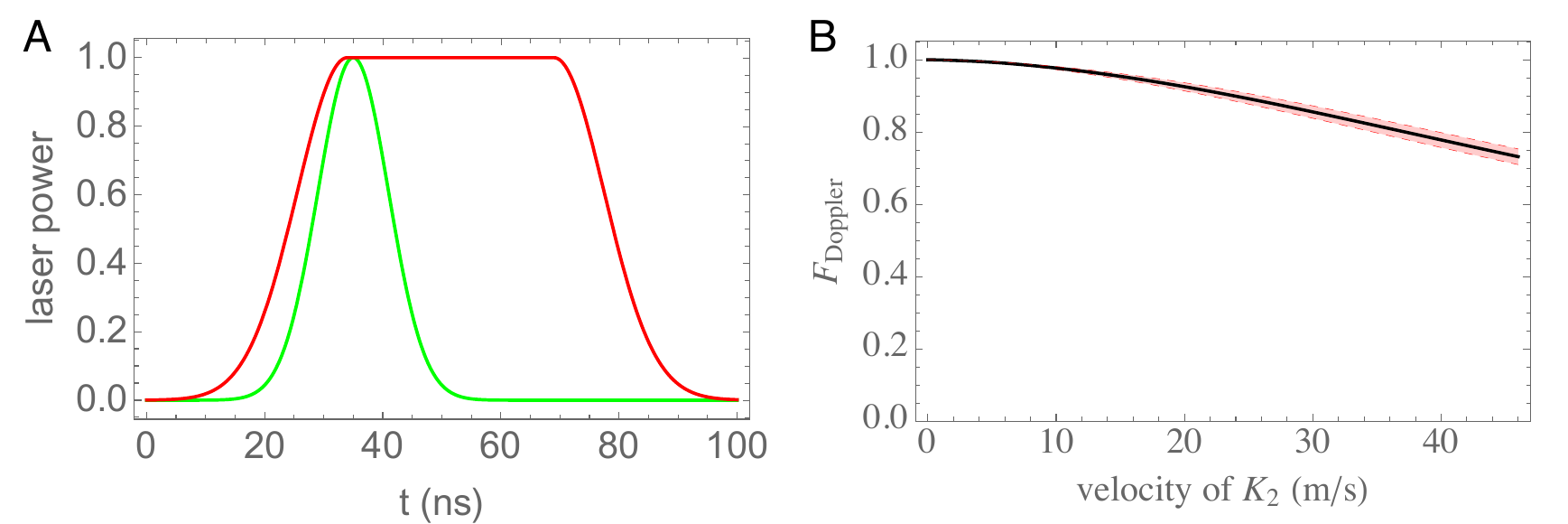}
    \caption{\normalsize\textbf{Modeling the Doppler effect for product sampling.} (A) Timing profiles of our REMPI laser pulses. (B) The Doppler factor $F_{\text{Doppler}}(v)$ versus the velocity of the K$_2$ product. The lower and upper bounds correspond to the situations with the peak value of $\Gamma_{\text{ion}}/2\pi$ at 6 and 14 MHz, respectively. }
\label{figDoppler}
\end{figure}

To characterize the Doppler effect on the product sampling, we use a method based on density matrix equations to analyze the dynamics of the REMPI process ~\cite{dixit1988resonant}. The $1+1'$ REMPI technique used here consists  of  an  initial  single-photon  bound-to-bound  transition  from  the  electronic  and  vibrational  ground-state $X^1\Sigma^+_g(v=  0,N)$  to  an  electronically  excited  intermediate-state $B^1\Pi_u(v',N')$, followed  by  a  single-photon  bound-to-continuum  transition  that  ionizes  the molecules. We drive the bound-to-bound transition using a frequency-tunable laser operating around 648 nm for the detection of K$_2$ and 674 nm for Rb$_2$. The bound-to-continum transition is excited by a 532 nm pulsed laser for both product species. For the convenience of discussion, the ground-state is denoted by $|0\rangle$ and the intermediate-state is denoted by $|1\rangle$. The density matrix that describes the dynamics of the REMPI process can be written as
\begin{eqnarray}
\frac{\text{d}}{\text{d}t}\rho_{00}(t)&=&-\frac{i}{2}(\Omega_{01}\rho_{10}-c.c.), \label{eq-rho00}\\
\frac{\text{d}}{\text{d}t}\rho_{11}(t)&=&-(\Gamma_1+\Gamma_{\text{ion}})\rho_{11}+\frac{i}{2}(\Omega_{01}\rho_{10}-c.c.), \\
\frac{\text{d}}{\text{d}t}\rho_{10}(t)&=&-\frac{1}{2}(\Gamma_1+\Gamma_{\text{ion}})\rho_{10}+i\Delta_1\rho_{10}+\frac{i}{2}\Omega_{10}(\rho_{11}-\rho_{00}),\label{eq-rho11}
\end{eqnarray}
where $\Omega_{01}$ is the Rabi frequency of the bound-to-bound transition, $\Delta_1=\omega-\omega_0$ is the detuning,  $\Gamma_1$ is the spontaneous decay rate of $|1\rangle$, and $\Gamma_{\text{ion}}$ is the ionization rate of the bound-to-continuum transition. Here $\hbar\omega_0$ represents the resonant transition energy and $\hbar \omega$ is the photon energy.  To consider realistic timing profile of the REMPI laser pulses, the corresponding time-dependent rates $\Omega_{01}(t)$ and $\Gamma_{\text{ion}}(t)$ are used in the numerical calculation (Fig. \ref{figDoppler}A). The ionization probability can be extracted via $P_{\text{ion}}=\int  P_{\text{loss}}(t)\frac{\Gamma_{\text{ion}}(t)}{\Gamma_1+\Gamma_{\text{ion}}(t)}dt$, with $P_{\text{loss}}(t)=1-\rho_{00}(t)-\rho_{11}(t)$. To take into account the Doppler effect, $\omega_0$ is replaced by $\omega_0(1+v_z/c)$ and the ionization probability thus becomes $v_z$-dependent, where $v_z$ is the projection of the product's velocity onto the propagation direction of the REMPI beams and $c$ is the speed of light. The formula of $P_{\text{ion}}(v_z)$, after substituting the corresponding values of the parameters including $\Gamma_1$, $\Gamma_{\text{ion}}$, $\Omega_{01}$ and $\Delta_1$, applies to both K$_2$ and Rb$_2$ products. We use $P_{\text{ion}}^{\text{K}_2}(v_z)$ and $P_{\text{ion}}^{\text{Rb}_2}(v_z)$ to denote the ionization probabilities of the two species, respectively. 

For the coincidence detection scheme used in this work, the probability of simultaneously ionizing a correlated product pair is  $P_{\text{ion}}^{\text{coin}}(v_z)=P_{\text{ion}}^{\text{K}_2}(v_z)P^{\text{Rb}_2}_{\text{ion}}(-m_{\text{K}_2}v_z/m_{\text{Rb}_2})$, where $v_z$ here represents the velocity projection of K$_2$ on the REMPI beam direction and $-m_{\text{K}_2}v_z/m_{\text{Rb}_2}$ is that of Rb$_2$ obtained based on the momentum conservation. By averaging over all possible directions of the product's velocity, a factor that characterizes the Doppler effect on the measurement efficiency of a correlated product pair is derived to be $F_{\text{Doppler}}(v)=\frac{1}{2v}\int_{-v}^{v}dv_z P_{\text{ion}}^{\text{coin}}(v_z)/P_{\text{ion}}^{\text{coin}}(v_z=0)$, where $v$ is the magnitude of K$_2$'s velocity and determined by the quantum states of products and the reaction exoergicity. The factor $F_{\text{Doppler}}(v)$ is calculated by solving the differential equations (\ref{eq-rho00}-\ref{eq-rho11}) as a function of $v$, as shown in Fig. \ref{figDoppler}B. In the calculation, we used $\Gamma_1/2\pi=14$ MHz,  $\Delta_1=0$, a peak value of $\Omega_{01}(t)$ at $2\pi\times$58 MHz, which is estimated using the calculated transition dipole moment and the measured laser intensity, and a peak value of $\Gamma_{\text{ion}}(t)$ at $2\pi\times(10\pm 4)$ MHz estimated based on the measured ionization efficiency.

\subsection{\large Normalization against fluctuations in experimental conditions}

In a given experiment where we probe a state-pair $| N_{\textrm{K}_2}, N_{\textrm{Rb}_2} \rangle$, the number of detected coincidence counts, $n_\textrm{coin}$, is proportional to the product of the state-dependent ionization probabilities, $P_{\textrm{ion}}(N_{\textrm{K}_2})P_{\textrm{ion}}(N_{\textrm{Rb}_2})$. Across different experiments, changes in the power, detuning, and relative timing of the REMPI lasers introduce fluctuations to $P(N_{\textrm{K}_2})$ and $P(N_{\textrm{Rb}_2})$, resulting in a biased sampling of coincidence counts. To correct against such an effect, we note that, in the same experiment, the total number of K$_2^+$ and Rb$_2^+$ ion counts, $n_{\textrm{K}_2^+}$ and $n_{\textrm{Rb}_2^+}$, are proportional to $P(N_{\textrm{K}_2})$ and $P(N_{\textrm{Rb}_2})$ respectively, and thus experience the same fluctuations. This allows us to construct a normalization factor as
\begin{equation}
    S(N_{\textrm{K}_2}, N_{\textrm{Rb}_2}) = \frac{n_{\textrm{K}_2^+}(N_{\textrm{K}_2}, N_{\textrm{Rb}_2})n_{N_{\textrm{Rb}_2}^+}(N_{\textrm{K}_2},N_{\textrm{Rb}_2})}{\langle{n_{\textrm{K}_2^+}(N_{\textrm{K}_2}' = N_{\textrm{K}_2},N_{\textrm{Rb}_2}')}\rangle  \langle{n_{\textrm{Rb}_2^+}(N_{\textrm{K}_2}',N_{\textrm{Rb}_2}' = N_{\textrm{Rb}_2})}\rangle},
\end{equation}
where $\langle{n_{\textrm{K}_2^+}(N_{\textrm{K}_2}' = N_{\textrm{K}_2},N_{\textrm{Rb}_2}')}\rangle$ is the average value of the number of $\textrm{K}_2^+$ ions from all experiments that share a common value for $N_{\textrm{K}_2}$, and $\langle{n_{\textrm{Rb}_2^+}(N_{\textrm{K}_2}',N_{\textrm{Rb}_2}' = N_{\textrm{Rb}_2})}\rangle$ is the average value of the number of $\textrm{Rb}_2^+$ ions from all experiments that share a common value for $N_{\textrm{Rb}_2}$.
We assess the effectiveness of this method of normalization by examining the relative differences in the coincidence counts obtained from two separate experiments on the same state-pairs, before and after applying the $S$ factor. From 11 different state-pairs where repeated data are available, we find that the average fluctuation in $n_\textrm{coin}$ decreases from $\alpha = 0.11$ to $\alpha = 0.054$ after normalization.
Since this method of normalization is somewhat \textit{ad hoc}, we retain the conservative value of $\alpha = 0.11$ for purposes of estimating experimental errors (section S5) and hypothesis testing (section S7) despite applying this normalization to all measured data.

\bigskip
\bigskip

Using the factors $F_{\textrm{geometry}}$, $F_{\textrm{Doppler}}$, and $S$ from the above derivations, as well as the number of cycles associated with each experiment, $C_{\textrm{exp}}$, we arrive at an overall normalization factor
\begin{equation}
    G(N_{\textrm{K}_2}, N_{\textrm{Rb}_2}) = \frac{C_{\textrm{exp}}(N_{\textrm{K}_2}, N_{\textrm{Rb}_2}) F_{\textrm{geometry}}(N_{\textrm{K}_2}, N_{\textrm{Rb}_2}) F_{\textrm{Doppler}}(N_{\textrm{K}_2}, N_{\textrm{Rb}_2}) S(N_{\textrm{K}_2}, N_{\textrm{Rb}_2})}{\sum_{\{N_{\textrm{K}_2}', N_{\textrm{Rb}_2}'\}} C_{\textrm{exp}}(N_{\textrm{K}_2}', N_{\textrm{Rb}_2}') F_{\textrm{geometry}}(N_{\textrm{K}_2}', N_{\textrm{Rb}_2}') F_{\textrm{Doppler}}(N_{\textrm{K}_2}', N_{\textrm{Rb}_2}') S(N_{\textrm{K}_2}', N_{\textrm{Rb}_2}')}.
\end{equation}
Here, the sum in the denominator is carried out over the data sets for all state-pairs that we probed in this study, including 57 allowed pairs ($\mathcal{S}$) and two forbidden ones ($| 10, 13 \rangle$ and $| 10, 15 \rangle$). 
Applying this overall factor to all detected coincidence counts, we obtain the normalized coincidence counts as $\mathcal{N}_{\textrm{coin}}(N_{\textrm{K}_2}, N_{\textrm{Rb}_2}) = n_{\textrm{coin}}(N_{\textrm{K}_2}, N_{\textrm{Rb}_2})/G(N_{\textrm{K}_2}, N_{\textrm{Rb}_2})$.
Note that due to the normalizations, $\mathcal{N}_{\textrm{coin}}$ can take on non-integer values. The scattering probabilities (Fig. 3) then derive from the normalized coincidence counts according to $P^{\textrm{meas}}_{\textrm{sc}}(N_{\textrm{K}_2}, N_{\textrm{Rb}_2}) = \mathcal{N}_{N_{\textrm{K}_2}, N_{\textrm{Rb}_2}}/\sum_\mathcal{S} \mathcal{N}_{N_{\textrm{K}_2}, N_{\textrm{Rb}_2}}$.

\section{\large~Likelihood ratio test for the statistical model} \label{likelihood ratio test}

The degree to which the measured product state distribution ($P_{\textrm{sc}}^{\textrm{meas}}$) agrees with the state-counting model ($P_{\textrm{sc}}^0$) is quantified using the likelihood ratio test ~\cite{wasserman2013all}. Formally, we test the null hypothesis $H_0: \pmb{\theta} \in \pmb{\Theta_0}$, where $\pmb{\theta} = \{ \mu_t(N_{\textrm{K}_2}, N_{\textrm{Rb}_2})\}_{\mathcal{S}}$ represents the set of mean coincidence counts given the true product state distribution, $\pmb{\Theta_0} = \{ \mu_0(N_{\textrm{K}_2}, N_{\textrm{Rb}_2})\}_{\mathcal{S}}$ represents the set of mean coincidence counts given the statistical model, and $\mathcal{S}$ is the entire set of allowed state-pairs. The values for $\mu_0$ depend both on the model and the experimental biases in our sampling of coincidence counts, and is expressed as
\begin{equation} \label{Eq: mean counts}
    \mu_0(N_{\textrm{K}_2}, N_{\textrm{Rb}_2}) =     \frac{P_{\textrm{sc}}^0(N_{\textrm{K}_2}, N_{\textrm{Rb}_2}) G(N_{\textrm{K}_2}, N_{\textrm{Rb}_2})}{\sum_{\mathcal{S}} P_{\textrm{sc}}^0(N_{\textrm{K}_2}, N_{\textrm{Rb}_2}) G(N_{\textrm{K}_2}, N_{\textrm{Rb}_2})} n_{\textrm{coin}}^{\textrm{tot}},
\end{equation}
where $G$ represents the overall normalization factor (section \ref{section: normalization of coincidence counts}), and $n_{\textrm{coin}}^{\textrm{tot}} = \sum_{\mathcal{S}} n_{\textrm{coin}}(N_{\textrm{K}_2}, N_{\textrm{Rb}_2})$ represents the sum of the set of all measured coincidence counts $\{ n_{\textrm{coin}}(N_{\textrm{K}_2}, N_{\textrm{Rb}_2})\}_{\mathcal{S}}$.

The likelihood for $H_0$ is given by
\begin{equation} \label{Eq: likelihood}
    \mathcal{L}_\mathcal{S}(\hat{\pmb{\theta_0}}) = \prod_{\mathcal{S}} p_c\left[n_{\textrm{coin}}(N_{\textrm{K}_2}, N_{\textrm{Rb}_2}), \mu_0(N_{\textrm{K}_2}, N_{\textrm{Rb}_2})\right].
\end{equation}
Here, $\hat{\pmb{\theta_0}}$ is the maximum likelihood estimate (MLE) when $\pmb{\theta}$ is restricted to lie in $\pmb{\Theta_0}$, and $p_c(n_{\textrm{coin}}, \mu_{0})$ is the probability of observing a particular count $n_{\textrm{coin}}$ given a mean count of $\mu_{0}$.
Since the accumulation of coincidence counts for each state-pair is a constant-rate process, we model $p_c$ as a Poisson distribution, but with an uncertainty in its mean introduced by experimental fluctuations captured by a Gaussian function. It is expressed as
\begin{equation}
    p_c(x, \mu,\alpha) = \left[ f(x) \ast g(\alpha) \right](\mu) = \int_{0}^{\infty} \left( \frac{e^{-m} m^x }{x !} \right) \left( \frac{1}{\alpha \mu \sqrt{2\pi}} e^{-\frac{1}{2} \left(\frac{m - \mu}{\alpha \mu} \right)^2}  \right) \textrm{d}m.
\end{equation}
Here, $f(x,m) = e^{-m} m^x / x! $ represents a Poisson distribution with mean $m$, and $g(m,\mu,\alpha) = \left( \alpha \mu \sqrt{2\pi} \right)^{-1} \exp\left[- (m - \mu)^2/(\alpha \mu)^2/2 \right] $ represents a normal distribution with mean $\mu$ and standard deviation $\alpha$. Note that $\alpha$ characterizes the relative fluctuation of the mean, and has an empirically-determined value of $ \alpha_0 = 0.11 $ for our measurements (section \ref{section: normalization of coincidence counts}).
Using the form for the likelihood function in Eq.\ref{Eq: likelihood}, we calculate the likelihood for the MLE of the parameters, $\pmb{\hat{\theta}}$, to be
\begin{equation} \label{Eq: likelihood MLE}
    \mathcal{L}_\mathcal{S}(\pmb{\hat{\theta}}) = \prod_{\mathcal{S}} p_c\left[n_{\textrm{coin}}(N_{\textrm{K}_2}, N_{\textrm{Rb}_2}), n_{\textrm{coin}}(N_{\textrm{K}_2}, N_{\textrm{Rb}_2})\right].
\end{equation}
Given $\mathcal{L}_{\mathcal{S}}(\hat{\pmb{\theta_0}})$ and $ \mathcal{L}_{\mathcal{S}}(\pmb{\hat{\theta}})$, we can calculate the likelihood ratio statistic as
\begin{equation} \label{Eq: lambda}
    \lambda_\mathcal{S} = 2 \log \left( \frac{\mathcal{L}_\mathcal{S}(\pmb{\hat{\theta}})}{\mathcal{L}_\mathcal{S}(\pmb{\hat{\theta}_0})} \right).
\end{equation}
Within the framework of the likelihood ratio test, the $p$-value for $H_0$ is
\begin{equation} \label{Eq: p-value}
    p_{\mathcal{S}} = \mathbb{P}(\chi^2_k > \lambda_{\mathcal{S}}),
\end{equation}
where $\chi^2_k$ is the chi-square distribution with $k$ degrees of freedoms, and $k$ is the length of $\mathcal{S}$, which is determined to be 57 in our study. We use $p_{\mathcal{S}} < 0.001$ as a threshold for rejecting $H_0$.

Using the values for $\{ n_{\textrm{coin}}(N_{\textrm{K}_2}, N_{\textrm{Rb}_2})\}_{\mathcal{S}}$ and $\{ \mu_0(N_{\textrm{K}_2}, N_{\textrm{Rb}_2})\}_{\mathcal{S}}$ obtained from this work, we find $\lambda_\mathcal{S} = 276 $ and $p_{\mathcal{S}} \sim 10^{-30}$ for $\alpha_0 = 0.11$. This means that $H_0$ should be rejected, and the outcome of our measurements is inconsistent with the state-counting model based on statistical theory.

On the other hand, we can test whether the model describes a subset of our measurements. To this end, we successively remove, from the above analysis, the state-pair that displays the largest deviation from the model prediction, and recalculate the $p$-value for $H_0$. To quantify the degree of deviation for each state-pair, we calculate its state-specific likelihood ratio statistic
\begin{equation}
    \lambda_{N_{\textrm{K}_2}, N_{\textrm{Rb}_2}} = 2 \log\left( \frac{p_c\left[n_{\textrm{coin}}(N_{\textrm{K}_2}, N_{\textrm{Rb}_2}), n_{\textrm{coin}}(N_{\textrm{K}_2}, N_{\textrm{Rb}_2})\right]}{p_c\left[n_{\textrm{coin}}(N_{\textrm{K}_2}, N_{\textrm{Rb}_2}), \mu_0(N_{\textrm{K}_2}, N_{\textrm{Rb}_2})\right]}\right)
\end{equation}
as well as the associated $p$-value, $p_{N_{\textrm{K}_2}, N_{\textrm{Rb}_2}} = \mathbb{P}(\chi^2_1 > \lambda_{N_{\textrm{K}_2}, N_{\textrm{Rb}_2}})$. The results are displayed in Tab. \ref{tab:productStates_exp}, in cases where we take the lower and upper bounds for $\alpha_0$, respectively. The state-pair that deviate the most from the state-counting model is $| 12, 7\rangle$ ($p = 10^{-66} \sim 10^{-18}$), which we have identified to be a near-threshold state-pair for which the long-range centrifugal barriers strongly suppresses product formation. Other state-pairs that strongly deviate include the ones with low translational energies, \textit{e.g.} $| 12, 5 \rangle$ ($p = 10^{-12} \sim 10^{-6}$) and $| 8, 15 \rangle$ ($p = 10^{-26} \sim 10^{-11}$).

Let $\mathcal{S}^{(j)} = \mathcal{S} - \{ \mathcal{S}_1,...,\mathcal{S}_j \}$ denote the reduced set for which the first $j$ members of $\mathcal{S}$ with the smallest state-specific $p$-values are removed. Using Eq. \ref{Eq: mean counts} -- \ref{Eq: p-value}, we calculate the $p$-values for the reduced sets, $p_{\mathcal{S}^{(j)}}$, for $j = 0,1,...,57$.
The results are shown in Fig. \ref{figProdStateDist}I. Here, we observe that the $p$-value for $H_0$ increases monotonically as we successively remove the largest outliers, and increases above the 0.001 threshold for rejecting $H_0$ after 7 state-pairs are removed. In other words, for a subset that contains the majority (50) of the allowed state-pairs, we find that the measured outcome to be consistent with the state-counting statistical model.

\section{\large~Distribution of product translational energy} \label{section: prod dist vs TE}

\begin{figure}[t!]
\centering
\includegraphics[width=1.00\textwidth]{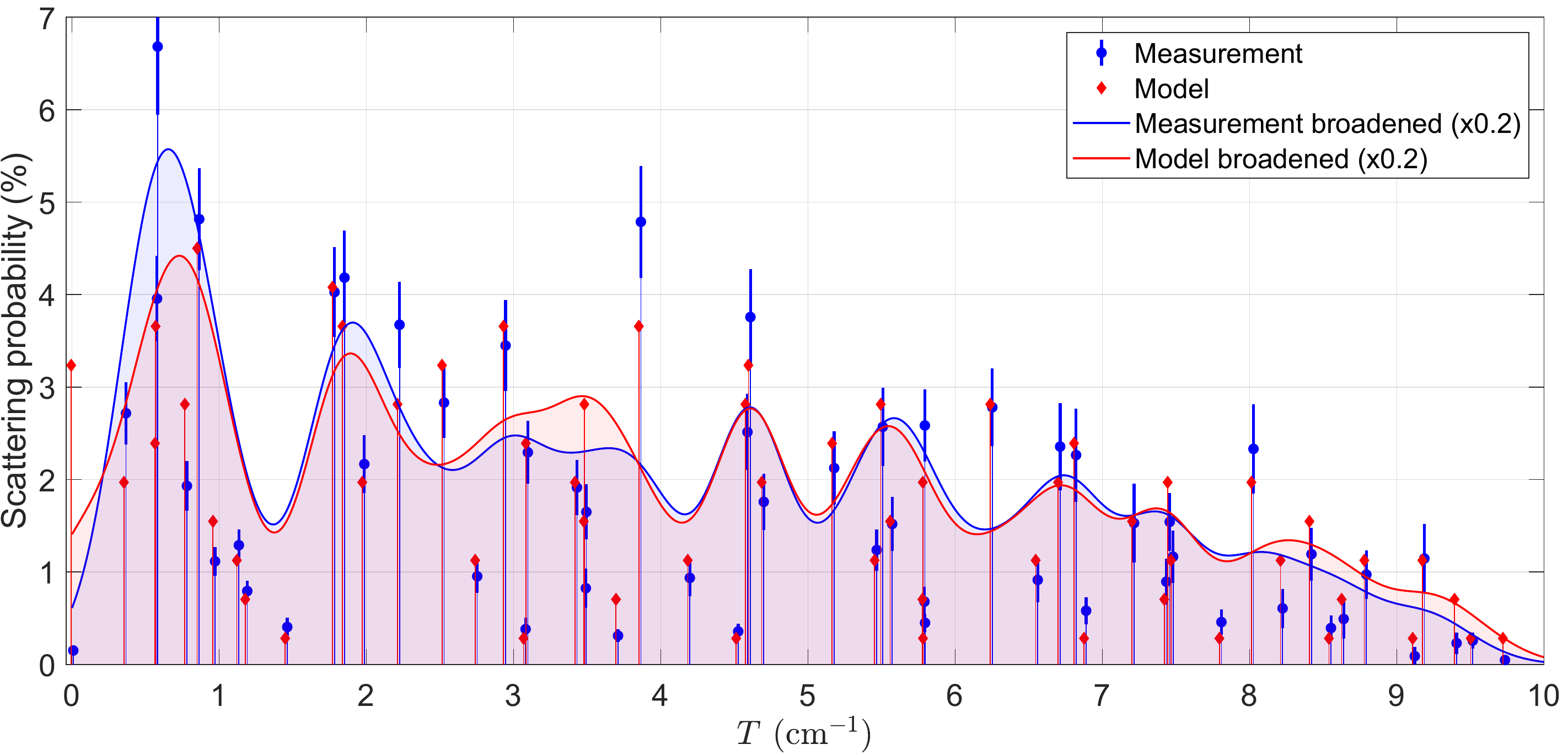}
    \caption{\normalsize\textbf{Distribution of product translational energy}. The measured (blue circle) and predicted (red diamond) scattering probabilities for all allowed state-pairs are plotted versus their translational energies. The two sets of data are offset horizontally by 0.014 cm$^{-1}$ for clarity. To aid in the identification of systematic deviations, we multiply each scattering probability by a normalized Gaussian function with a 1$\sigma$ width of 0.25 cm$^{-1}$, and sum them up to construct broadened distributions as shown by the blue and red curves. These curves are scaled by a factor of 0.2 for convenience.}
\label{figDistVsTE}
\end{figure}

In a 2012 review article~\cite{nesbitt2012toward}, Nesbitt speculated that energy deposition into product translation should be dynamically disfavored for ultracold complex-forming reactions such as KRb + KRb. The basis for this speculation is the observed propensity for complexes bound by van dar Waals forces to strongly favor the formation of low translational energy products upon their dissociation. Here, we examine whether such a propensity exist in our result by displaying the measured and predicted scattering probabilities for all allowed product state-pairs as functions of their translational energy, and searching for any systematic trends in how they deviate from each other (Fig. \ref{figDistVsTE}). To this end, we construct ``blurred" distributions by apply a Gaussian broadening of 0.25 cm$^{-1}$ 1$\sigma$ width to the measured and predicted amplitude of each state-pair. Comparing the two blurred distributions, we do not observe any strong monotonic trends in their difference.

\section{\large~Product escape probabilities}

In this section, we calculate the probability for products to escape from the complex and into each allowed state-pair. For this purpose, we consider the microscopically reverse process of product capture into the complex, and the associated capture probability $\mathcal{C}(N_{\textrm{K}_2}, N_{\textrm{Rb}_2})$.
The details of the implementation for diatom-diatom systems have been published in Ref.~\cite{yang2020statistical}. In brief, the time-independent Schrödinger equation was solved in diatom-diatom Jacobi coordinates ($R$,$r_1$,$r_2$,$\theta_1$,$\theta_2$,$\phi$) using the log-derivative method~\cite{johnson1973multichannel,manolopoulos1986improved} with the Wentzel-Kramers-Brillouin (WKB) approximation~\cite{johnson1973generalized} to damp the wavefunction within the capture radius. The interaction potential was determined using a similar method as in Ref. ~\cite{yang2020statistical}. 
In the calculations, only the $p$-wave ($L_{\textrm{reac}}=1$) and $\epsilon = -1$ were considered, given the fermionic nature of the KRb reactants. 3/3/15/15/20 points were used for the $r_1$/$r_2$/$\theta_1$/$\theta_2$/$\phi$ quadratures. The log-derivative propagation steps were chosen as $\Delta R=0.05$ $a_0$ for $R\in[35.0, 80.0]$ $a_0$,  $\Delta R=0.5$ $a_0$ for $R\in[80.0, 200.0]$ $a_0$, and $\Delta R=1.00$ $a_0$ for $R\in[200.0, 800.0]$ $a_0$, respectively. The number of rotational bases is chosen to be $N_{\textrm{K}_2}^{\textrm{max}} = 20$ and $N_{\textrm{Rb}_2}^{\textrm{max}} = 30$. The effective potentials for the calculation are defined in the manner of adiabatic channel potential energy, as
\begin{equation}
    V_{\xi}^{\textrm{eff}}(R) = W_{\xi, \xi}(R) + \frac{L_{\textrm{prod}}(L_{\textrm{prod}}+1)}{2 \mu_{\textrm{K}_2,\textrm{Rb}_2} R^2},
\end{equation}
Here, $\xi = \{N_{\textrm{K}_2},N_{\textrm{Rb}_2},N_{\textrm{prod}},L_{\textrm{prod}}\}$ is the set of product quantum numbers defined in section S4, $W_{\xi, \xi}$ is the diagonal element of interaction potential matrix in the space-fixed frame, which can be calculated by an orthogonal transformation from the body-fixed counterpart.
The results show that all allowed state-pairs besides $|12, 7\rangle$ have effectively unit probabilities ($\mathcal{C} > 0.999$) to escape the complex, indicating that product formation in these state-pairs are not hindered by any barriers or bottlenecks.

\bibliography{refs}

\end{document}